\documentstyle[aps,epsf]{revtex}

\begin{document}
\draft

\author{
K.A. Kikoin$^1$, M.N. Kiselev$^{2,3}$, A.S. Mishchenko$^2$, 
and A. de Visser$^4$}

\address{
$^1$ {\it Ben-Gurion University of the Negev, Beer-Sheva 84105, Israel}\\
$^2$ {\it Russian Research Center "Kurchatov Institute", Moscow 123182,
Russia}\\
$^3$ {\it Laboratoire Leon Brillouin, CE-Saclay 91191 Gif-sur-Yvette Cedex,
France}\\
$^4$ {\it Van der Waals-Zeeman Institute, University of Amsterdam,\\
Valckenierstraat 65, 1018 XE Amsterdam, The Netherlands}\\
}
\date{\today}

\title{
THERMODYNAMICS OF CeNiSn AT LOW TEMPERATURES \\
AND IN WEAK MAGNETIC FIELDS}
\maketitle

\begin{abstract}
Detailed experimental and theoretical studies of the low-temperature
specific heat, magnetic susceptibility, thermal expansion and
magnetostriction of the orthorhombic compound CeNiSn are presented. All
anomalies observed in the thermodynamic and magnetic properties of CeNiSn
are explained in a framework of a model of metallic Kondo lattice with well
developed spin-liquid-type excitations. The pseudogap behavior of these
excitations appears due to interplay between spin-fermions and soft
crystal-field (CF) states. The thermodynamic relations for the spin liquid
are derived. Together with the explanation of inelastic neutron scattering
spectra given earlier within a same approach these studies of the
low-temperature thermodynamics and magnetic response give a consistent
description of the nature of anomalies in the low-temperature thermodynamics
of perfect and imperfect CeNiSn crystals.
\\
\mbox{} 
\end{abstract}

\section{Introduction}

The orthorhombic compounds CeNiSn and CeRhSb are known as Kondo lattice
systems with peculiar thermodynamic and magnetic properties. Unusual
features are observed at low temperatures $T<T^{*}$ in the specific heat,
the thermal expansion coefficient, the magnetic susceptibility, the
magnetostriction, and the NMR relaxation rate (see \cite{Taka93,Taka95} for
a review of early data). The characteristic temperature $T^{*}$ is $\sim 10K$
for both systems. It should be emphasized that this temperature is much less
than the Kondo temperature $T_{K}$ estimated by standard methods, e.g.,
extracted from the logarithmic high-temperature dependence of the electrical
resistivity. In the early measurements the electrical resistivity showed an
upturn at low temperatures in the temperature region $T<T^{*}$. This upturn
was interpreted as the indication to a non-metallic ground state of these
systems, and the energy gap in the heavy electron spectrum was claimed to be
responsible for the peculiar behavior of CeNiSn and CeRhSb. These materials
together with the cubic Ce- and U-based compounds, ${\rm Ce_{3}Sb_{4}Pt_{3}}$
or ${\rm U_{3}Bi_{4}Pt_{3}}$ were classified as ''Kondo insulators'' \cite    
{Aeppli92}.

Later on it turned out that significant differences exist between the
real-gap cubic semiconductors and the orthorhombic CeNiSn family (see \cite
{Taka98} for a review). Most striking was the observation that the CeNiSn
single crystals of good quality show metallic character of the resistivity 
\cite{Taka96} at very low temperatures, and such behavior seems to be
incompatible with the idea of a gap or pseudogap in the electron spectrum.
Comparing the metallic behavior of electron transport with the anomalous
low-temperature thermodynamics, one could suspect that the electronic
spectrum with the pseudogap used in early phenomenological theories hardly
can be responsible for all low-$T$ peculiarities observed in the physical
properties of CeNiSn and CeRhSb. Meaningful, e.g., is the fact that the
unusual temperature dependence of the NMR relaxation rate $1/T_{1}\sim T^{3}$
which was explained by the V-shape form of the density of electron states
around the chemical potential at the bottom of the pseudogap \cite{Taka93}
is observed exactly in the same temperature interval where the conventional
Fermi-liquid-type $T^{2}$ law is seen for the electrical resistivity \cite
{Taka96}. At $T<1$K the relaxation rate obeys the linear-$T$ Korringa law
characteristic for fermions with constant density of states \cite{Nakam95}.
One more striking feature of the low-energy excitations in CeNiSn is the
extremely complicated $({\bf Q},\omega )$-dependent structure of the
inelastic magnetic scattering spectra that was observed in the same
temperature region $T<T^{*}$ \cite{Mason92,Sato95}. In gross features these
unusual spectra also can be interpreted in terms of a pseudogap in the
spin-excitation spectrum \cite{Ikeda96}, although this phenomenology seems
to be too simplistic to explain numerous details of the highly anisotropic
neutron scattering cross section.

The theoretical approaches to the problem either implement the idea of a
Kondo insulator with all its shortcomings, or try to offer alternative
mechanisms which are based on a metallic type of electron spectra and seek
the explanation of low-temperature thermodynamics and magnetic response in
the unusual properties of the magnetic excitations. In the first case the
starting point of the theory is the mean-field slave-boson approximation to
the Anderson lattice \cite{Riseb92,Ono94}. The latest version of mean-field
hybridization theory \cite{Ikeda96} refers to the actual symmetry of
f-electron states in the orthorhombic crystal. Since this procedure implies
strong coupling of spin and charge degrees of freedom, the gap (or
pseudogap) in the excitation spectrum necessarily means a semiconductor or
semimetallic type of electrical resistivity which, apparently, contradicts
the experimental data mentioned above.

An alternative approach was offered in Ref.\cite{KKP93}. In this theory new
characteristic features with an energy scale of $T\ll T_{K}$ appear in the
spectrum of the spin excitations due to the interplay between the non-local
(spin-liquid) excitations characterized by the energy scale of $T_{K}$ and
the single-site crystal field (CF) excitations with the energy $\Delta
_{CF}<T_{K}$. Within this model the semi-quantitative description of the
low-energy specific heat and the thermal expansion coefficient was given in
Ref.\cite{KVBT94}. The CF levels are not seen directly neither in CeNiSn,
nor in CeRhSb \cite{Taka93}, and this result indicates that these local
excitations are ''dissolved'' in the continuum of low-energy excitations of
the Kondo lattice. However, the indirect estimate of the magnitude of
crystal field created by the Ni ions on the Ce site \cite{Aleks94} confirmed
the validity of the inequality $\Delta _{CF}<T_{K}$.

Basing on the available experimental data related to the structure of
magnetic excitations in CeNiSn, the quantitative theory of interplay between
heavy fermions and CF excitations in CeNiSn was offered in Ref. \cite{KAKM97}%
. The theory involves the idea of spin-liquid excitations of resonating
valence bond (RVB) type which transform at low temperature into well defined
fermions with their own dispersion predetermined by dispersion of RKKY
exchange in the Brillouin zone \cite{KKM94}. As a result of the interplay
between these excitations and soft CF states, the spectrum of spin-fermions
in the low-symmetry lattice of CeNiSn transforms in such a way that the a
deep minimum appears in the spin density of states (DOS) in the vicinity of
the spinon Fermi level. Since the spin-liquid excitations are decoupled from
the charged Fermi-liquid excitations in the conduction band, the gap in the
spin DOS does not imply a corresponding gap in the electron DOS, and the
system possesses metallic conductivity whereas the spin excitations are
responsible for the thermal properties. It was shown in Ref. \cite{KAKM97}
that the inelastic transitions between the spinon states in the Brillouin
zone are responsible for the complicated picture of inelastic magnetic
neutron scattering. The successful attempt of the quantitative description
of the magnetic scattering function $S({\bf Q},\hbar \omega )$, gives us
strong arguments in favor of the existence of spin-liquid correlations in
CeNiSn and related materials. Moreover, the fitting of the theoretical
spectra to the experimental $S({\bf Q},\hbar \omega )$ provided us the
values of the model parameters. With these data at hand we are able to give
a quantitative description of the temperature dependence of various
thermodynamical functions on the assumption \cite{KKP93,KVBT94} that the
spin-liquid-type excitations give the main contribution to the low-$T$
thermodynamics. Thus, a unified description of the low-temperature and
low-energy properties of the orthorhombic CeNiSn family becomes possible.

The main purpose of the present paper is to give a detailed experimental and
theoretical picture of the low-temperature specific heat, magnetic
susceptibility, thermal expansion and magnetostriction coefficients in the
CeNiSn family. This description should be consistent with the picture of
magnetic excitations, as given by the inelastic neutron scattering
experiments. Some of the experimental data for CeNiSn and CeRhSb were
published in Refs. \cite{KVBT94,Nolt95}. The first attempts of describing
the inelastic magnetic spectra, low-temperature specific heat, and thermal
expansion of these systems by using the same model \cite{KAKM97,KKM96}
demonstrated the validity of the spin-liquid description.

\section{Experimental}

The magnetostriction of single-crystalline CeNiSn was measured in magnetic
fields up to 8 T at selected temperatures of 0.5, 1.4 and 4.3 K. The
magnetostriction is defined by $\lambda=(L(B)-L(0))/L(0)$, where $L(0)$ is
the length of the specimen along a certain crystallographic direction in
zero magnetic field. The magnetic field was always applied along the
orthorhombic $a$-axis, while $\lambda$ was measured along the field
direction ($\lambda_a$) and perpendicular to the field direction along the $%
b $- ($\lambda_b$) and the $c$-axis ($\lambda_c$). The volume
magnetostriction is defined by $\lambda_v= \lambda_a+\lambda_b+\lambda_c$
for a fixed field direction. The field was applied along the $a$-axis
because this is the easy axis for magnetization: the low-temperature
susceptibility $\chi_a$ is about a factor two larger than $\chi_b$ and $%
\chi_c$, and, moreover $\chi_a$ exhibits a pronounced maximum at 12 K.

The experiments were carried out on a Czochralski grown single-crystalline
sample. The sample was shaped by means of spark erosion into a cube with
edges along the principal axes of the orthorhombic unit cell ($a\times
b\times c\approx 2\times 2\times 2$mm$^{3}$). The magnetostriction was
measured using a sensitive parallel-plate capacitance cell machined of
oxygen free high-conductivity copper. The magnetostriction cell was fixed
to the cold plate of a $^{3}$He-insert, which is operated with an adsorption
pump. The $^{3}$He-insert could be placed in a superconducting solenoid with 
$B_{max}=8$ T. The magnetostriction was measured by recording the
capacitance, while slowly sweeping the field. Temperatures were stabilized
by regulating on a field- insensitive RuO$_{2}$ chip-resistor which served
as thermometer.

The experimental results are shown in Fig. 1a-c, while the coefficients of
magnetostriction $\lambda _{i}^{\prime }=L^{-1}dL/dB$, obtained by
differentiating the data of Fig. 1 with respect to the field, are shown in
Fig. 2a-c. At the highest temperature, $T=4.3$ K, $\lambda _{i}$ ($i=a,b,c$)
is a monotonous function of the field. The crystal expands in $a-b$ plane
and shrinks along $c$-axis when $B\parallel a$. The magnetostriction is
anomalous in the sense that the curves $\lambda _{i}^{\prime }(B)$ deviate
from the standard linear behaviour for paramagnetic systems. At lower
temperatures this anomalous behaviour becomes stronger and $dL_{i}/dB$
change their signs with increasing field. For instance, at $T=0.5$ K, the $%
a-b$ plane shrinks till $\sim $ 4.5 T and the $c$-axis expands till $\sim $
7 T. The anomalous behavior is also seen in the volume magnetostriction as a
negative contribution at low temperatures, although is not very pronounced.

The magnetostriction data are in good agreement with previous thermal
expansion measurements in zero and applied magnetic fields ($B\parallel a$)
of 4 and 8 T, taken on the same single-crystalline specimen \cite{KVBT94}.
Strong anisotropy is observed in the linear expansion ($\alpha
_{i}=L^{-1}(dL_{i}/dT)$): the dependence $\alpha _{c}(T)$ is anomalous with
respect to $\alpha _{a,b}(T)$. In magnetic field a sign reversal takes place
at low temperatures ($T<3$ K at 8 T ): $\alpha _{c}(T)$ becomes negative,
while $\alpha _{a}$ and $\alpha _{b}$ become positive. The $\alpha _{i}(T)$%
-curves show several anomalies, but the coefficient of volume expansion, $%
\alpha _{v}=\alpha _{a}+\alpha _{b}+\alpha _{c}$, is monotonous. Our
magnetostriction data are also in excellent agreement with the data reported
in Ref. \cite{Holt96} in the temperature range 0.1- 4.2 K and field range up
to 20 T.

It is known that the reversible volume magnetostriction is thermodynamically
equivalent to the strain dependence of the magnetic susceptibility $\chi
(B,T)$ \cite{ChanFaw71} 
\begin{equation}
\lambda _{V}^{\prime }(B,T)=\kappa _{T}B\left( \frac{\partial \chi (B,T)}{%
\partial \log V}\right) _{T,B}  \label{anne1}
\end{equation}
(the magnetic susceptibility is defined as $\chi (B,T)=M(B,T)/B$, where $%
M(B,T)$ is the magnetization). Therefore, one can extract from the
experimental result the field and temperature dependence of the magnetic
susceptibility volume derivative $(\partial \chi (B,T)/\partial \log
V)_{B,T} $. Introducing the doubly differential magnetostriction coefficient 
\begin{equation}
\lambda _{V}^{\prime \prime }(B,T)=B^{-1}(d\lambda _{V}/dB)_{V,T}
\label{anne2}
\end{equation}
one can express the logarithmic volume derivative of the magnetic
susceptibility as 
\begin{equation}
\left( \frac{\partial \chi (B,T}{\partial \log V}\right) _{V,T}=\frac{%
\lambda _{V}^{\prime \prime }(B,T)}{\kappa _{T}}\quad .\   \label{anne3}
\end{equation}
It is seen from Fig. 3 that the temperature and field dependence of the
volume derivative of the magnetic susceptibility deviates from the normal
temperature and field independent behaviour at low temperatures and in low
magnetic fields.

\section{Hamiltonian and the energy of the spin liquid}

The orthorhombic compounds CeNiSn and CeRhSb are usually classified as Kondo
lattices with moderately heavy fermion (HF) properties. The basic
Hamiltonian which describes the Ce-based HF systems is the Anderson lattice
Hamiltonian for the Ce$^{3+}(f^{1})$ ion hybridized with the conduction
electrons. In the Kondo lattice limit when the valence of the Ce ion is
close to integer, one deals with well localized f-electrons for which the
inequality $V_{k_{F}\Lambda }^{{\bf i}}\ll \epsilon _{F}-E_{\Gamma }$ is
believed to be valid (here $V_{k\Lambda }^{{\bf i}}$ is the hybridization
matrix element between the f-electron localized on a site ${\bf i}$ in a
state $|\Lambda \rangle =|\Gamma \nu \rangle $ with the energy $E_{\Gamma }$
of the f-electron in a crystal field and the partial component of the Bloch
wave $|k\Lambda \rangle $, $\nu $ is the row of the irreducible
representation $\Gamma $ of the crystal point group, $\epsilon _{F}$ is the
Fermi energy of conduction electrons). This hybridization integral is taken
in the Coqblin-Cornut (CC) approximation \cite{Corn72} which represents the
Bloch functions by their partial waves $c_{k\Lambda }^{\dagger }$, and takes
into account only the diagonal in $\Lambda $ hybridization matrix elements $%
V_{k\Lambda }^{{\bf i}}=\langle k\Lambda |V|{\bf i}\Lambda \rangle $. Then
the hybridization effects are reduced to exchange-like interaction between
the localized f-electrons and the conduction electrons with an effective
coupling constant 
\[
J_{{\bf i}}^{\Lambda \Lambda ^{\prime }}(k,k^{\prime })=V_{k\Lambda }^{{\bf i%
}*}V_{k^{\prime }\Lambda ^{\prime }}^{{\bf i}}/(\epsilon _{k}-E_{\Gamma }). 
\]

As was shown in \cite{KKP93,KAKM97}, the non-CC hybridization $\bar{V}%
_{k\Lambda }^{{\bf i}\Lambda ^{\prime }}=\langle {\bf i}\Lambda |V^{\prime
}|k\Lambda ^{\prime }\rangle $ is of crucial importance for the interplay
between the one-site crystal-field excitations and the non-local spin liquid
excitations (here $V^{\prime }$ is the component of the crystal field which
has a symmetry lower than that diagonalizing the f-electron energy terms $%
E_{\Gamma }$. Respectively, the non-CC effective exchange constant is
introduced as 
\[
\widetilde{J}_{{\bf i}}^{\Lambda \Lambda ^{\prime }}(k,k^{\prime })=\bar{V}%
_{k\Lambda }^{{\bf i}\Lambda ^{\prime }*}V_{k^{\prime }\Lambda ^{\prime }}^{%
{\bf i}}/(\epsilon _{k}-E_{\Gamma }). 
\]

In the case of completely suppressed charge fluctuations in the f-channel
the sf-exchange can be taken into account in the second order approximation,
and one comes to the effective RKKY-like Hamiltonian, where the f-electrons
are represented only by their spin degrees of freedom described by the
spin-fermion operators $f_{{\bf i}\Lambda }$. When the CF excitations are
involved, this Hamiltonian acquires the following form (detailed derivation
of $H^{s}$ can be found in Refs. \cite{KAKM97,coop85}): 
\begin{equation}
H^{s}=H_{f}+H_{h}+H_{RKKY}^{(c)}+H_{RKKY}^{(nc)}\;.  \label{1.900}
\end{equation}
Here 
\begin{equation}
H_{f}=\sum_{{\bf i},\Lambda }E_{\Gamma }|{\bf i}\Lambda \rangle \langle {\bf %
i}\Lambda |  \label{1.2}
\end{equation}
describes the Ce$(f^{1})$-ions on the lattice sites. 
\begin{equation}
H_{h}=\sum_{{\bf i}}\left[ \sum_{\Lambda \Lambda ^{\prime }}{\cal B}_{{\bf i}%
}^{\Lambda \Lambda }\delta _{\Lambda \Lambda ^{\prime }}f_{{\bf i}\Lambda
}^{\dagger }f_{{\bf i}\Lambda }\;+\;{\cal \widetilde{B}}_{{\bf i}}^{\Lambda
\Lambda ^{\prime }}f_{{\bf i}\Lambda }^{\dagger }f_{{\bf i}\Lambda ^{\prime
}}(1-\delta _{\Lambda \Lambda ^{\prime }})\right]  \label{1.61}
\end{equation}
corresponds to effective covalent contribution to the one-site CF splitting
due to virtual sf-transitions. Here 
\[
{\cal B}_{{\bf i}}^{\Lambda \Lambda }=-\sum_{{\bf k}}\frac{\bar{V}_{k\Lambda
}^{{\bf i}*}V_{k\Lambda }^{{\bf i}}}{\epsilon _{k}-E_{\Gamma }},\;\;\;{\cal 
\widetilde{B}}_{{\bf i}}^{\Lambda \Lambda ^{\prime }}=-\sum_{{\bf k}}\frac{%
\bar{V}_{k\Lambda }^{{\bf i}\Lambda ^{\prime }*}V_{k\Lambda }^{{\bf i}}}{%
\epsilon _{k}-E_{\Gamma }}\quad .~ 
\]
The effective exchange interaction mediated by conduction electrons is given
by the last two terms in the Hamiltonian (\ref{1.900}) 
\begin{equation}
H_{RKKY}^{(c)}=\sum_{{\bf ii^{\prime }}}^{{\bf i}\ne {\bf i}^{\prime
}}\sum_{\Lambda \Lambda ^{\prime }}{\cal I}_{{\bf ii^{\prime }}}^{\Lambda
\Lambda ^{\prime }}f_{{\bf i}\Lambda }^{\dagger }f_{{\bf i}\Lambda ^{\prime
}}f_{{\bf i^{\prime }}\Lambda ^{\prime }}^{\dagger }f_{{\bf i^{\prime }}%
\Lambda }\quad ,  \label{1.18}
\end{equation}
and the non-CC interaction is represented by the last term $H_{RKKY}^{(nc)}$
which is responsible for the interplay between the HF and CF excitations in
our model,

\begin{equation}
H_{RKKY}^{(nc)}=\sum_{{\bf ii^{\prime }}}\sum_{\Lambda \Lambda ^{\prime
}\Lambda ^{\prime \prime }}^{\Lambda \ne \Lambda ^{\prime \prime }}\left[ 
\bar{{\cal I}}_{{\bf ii^{\prime }}}^{\Lambda \Lambda ^{\prime }\Lambda
^{\prime \prime }\Lambda ^{\prime }}f_{{\bf i}\Lambda }^{\dagger }f_{{\bf i}%
\Lambda ^{\prime }}f_{{\bf i^{\prime }}\Lambda ^{\prime }}^{\dagger }f_{{\bf %
i^{\prime }}\Lambda ^{\prime \prime }}+H.c\right] \;.  \label{1.182}
\end{equation}
This is the lowest in $\bar{V}_{k}$ term among the non-CC indirect exchange
interactions which admix the excited CF states $\mid \Lambda \rangle =\mid
E\nu ^{\prime }\rangle $ to the ground state doublet $\mid \Lambda \rangle
=\mid G\nu \rangle $.

The uniform spin-liquid state in the Heisenberg-like Hamiltonians with
antiferromagnetic exchange constant is described by the free energy
expression 
\begin{equation}
{\cal F}=\beta ^{-1}\int_{0}^{\beta ^{-1}}{\cal E}(\beta ^{\prime })d\beta
^{\prime }-\beta ^{-1}{\cal S}_{\infty }  \label{11.2}
\end{equation}
where $\beta ^{-1}=k_{B}T$, ${\cal S}_{\infty }$ is the magnetic entropy at $%
T\rightarrow \infty $, and ${\cal E}$ is the average value of the
Hamiltonian expressed via the two-spinon correlator. In the case of the
isotropic Heisenberg Hamiltonian this average energy is given by 
\begin{equation}
{\cal E}=\sum_{ii^{\prime }}\frac{{\cal I}_{ii^{\prime }}}{2}\langle |\Delta
_{ii^{\prime }}|^{2}\rangle  \label{11.1}
\end{equation}
where 
\begin{equation}
\Delta _{ii^{\prime }}=\sum_{\alpha }f_{i\alpha }^{\dagger }f_{i^{\prime
}\alpha }  \label{11.3}
\end{equation}
is the non-local operator creating the RVB pair, $\alpha $ stands for the
''flavor'' (e.g., spin projection in case of pure spin states). After
Fourier transformation the average energy of the uniform spin liquid
acquires the form 
\begin{equation}
{\cal E}=\frac{{\cal I}}{2}\sum_{{\bf pq}}\sum_{\alpha \alpha ^{\prime
}}\varphi _{{\bf p-q}}\langle \Delta _{{\bf p}}\Delta _{{\bf q}}\rangle
\quad .  \label{11.4}
\end{equation}
Here $\varphi _{{\bf k}}=\sum_{n}\exp (-i{\bf kR}_{n})$ is the structure
factor for the exchange interaction.

Usually, in 3D Heisenberg lattices the spin liquid state has higher energy
than the AFM state \cite{Anders73}, and the standard mean field approach
predicts magnetic order at low temperatures. However, the mechanism
stabilizing the spin liquid state in Kondo lattices was proposed in Refs. 
\cite{KKM94,Cole89}. It was shown within the mean-field approximation that
the AFM phase can be suppressed by Kondo-type screening, provided ${\cal I}%
\sim k_{B}T_{K}$, and that the spin-liquid state which is not that sensitive
to Kondo scattering can be realized instead. Recently it was pointed out 
\cite{KKM97} that the influence of low-lying relaxation modes in the spin
system can transform the phase transition to the spin liquid state into a
crossover. The low-lying excitation mode (in particular, the soft CF
excitations) can play a similar role in stabilization of the spin liquid
state, and the thermodynamics in this case should be described by the
equation generalizing eq. (\ref{11.4}) for the case of CF excitations
admixed to the ground Kramers state of the rare-earth ion. The dynamical
correlation function $\langle \Delta _{{\bf p}}\Delta _{{\bf q}}\rangle
_{\omega }$ determines the frequency dependence of inelastic magnetic
neutron scattering \cite{KAKM97}, so the possibility opens for a unified
description of low-temperature thermodynamics and the low-energy spin
excitations.

To realize this possibility we first should derive the expression for the
free energy of the Kondo lattice described by the Hamiltonian $H^s$ (\ref
{1.900}). This means that we should find the energy ${\cal E}$ or,
eventually, to diagonalize the matrix 
\begin{equation}
{\sf M}={\sf H}^s-{\sf 1}\cdot {\sf E}  \label{11.5}
\end{equation}
in terms of the variables $\Delta$.

Having in mind the low symmetry of the CeNiSn lattice, we consider the
general case of the elementary cell {\bf l} containing several sublattices $%
\xi =1,...,L$ possessing the same point symmetry group as the Ce ion which
total magnetic moment is $J=5/2$. When diagonalizing $\langle H^{s}\rangle $
given by eq. (\ref{1.900}) in terms of spin liquid variables, we introduce a
single anomalous correlator $\Delta ^{G}=\langle f_{{\bf l^{\prime }}\xi
^{\prime }G}^{\dagger }f_{{\bf l}\xi G}\rangle $, which corresponds to the
ground state doublet $\Lambda =G$. In the course of the diagonalization
procedure it turns out that this parameter determines the dispersion of both
the lower and the higher branches of the excitation spectrum which arise due
to interplay between CF and HF excitations (see Appendix A).

We introduce the Fourier transformation 
\begin{equation}
f_{{\bf l}\xi \Lambda }^{\dagger }=N^{-1/2}\sum_{{\bf k}\nu }e^{i{\bf kl}%
}\Theta _\nu ^\Lambda (\xi ,{\bf k})f_{{\bf k}\nu }^{\dagger }  \label{a8}
\end{equation}
to the basis $\left\{ {\bf k}\nu \right\} $ which diagonalizes the
translationally invariant matrix ${\bf {\rm M}}$ (eq. \ref{11.5}) and the
averages $\langle f_{{\bf l^{\prime }}\xi ^{\prime }\Lambda }^{\dagger }f_{%
{\bf l}\xi \Lambda }\rangle $ ($\nu =1,...,(2J+1)L$). The eigenvectors $%
\Theta _\nu ^\Lambda $ are orthonormal, 
\begin{eqnarray}
\sum_\nu \Theta _\nu ^\Lambda (\xi ,{\bf k})[\Theta _\nu ^{\Lambda ^{\prime
}}(\xi ^{\prime },{\bf k})]^{*} &=&\delta _{\Lambda \Lambda ^{\prime
}}\delta _{\xi \xi ^{\prime }},  \nonumber \\
\sum_{\Lambda \xi }\Theta _\nu ^\Lambda (\xi ,{\bf k})[\Theta _{\nu ^{\prime
}}^\Lambda (\xi ,{\bf k})]^{*} &=&\delta _{\nu \nu ^{\prime }}\quad .
\label{a81}
\end{eqnarray}

As a result, the average energy becomes the functional of the ''occupation
numbers'' 
\begin{equation}
\langle f_{{\bf k}\nu }^{\dagger }f_{{\bf k}^{\prime }\nu ^{\prime }}\rangle
=n_{{\bf k}\nu }\delta _{{\bf kk}^{\prime }}\delta _{\nu \nu ^{\prime }}\;,
\label{a9}
\end{equation}
and the final equation for energy per Ce ion ${\cal E}(T,\{n_{{\bf k}}\})$
has the form 
\begin{equation}
{\cal E}(T,\{n_{{\bf k}}\})=\frac{\Delta _{CF}^{(0)}}{NL}\sum_{{\bf k}\nu
}n_{{\bf k}\nu }\Phi _{{\bf k}\nu }(\{n_{{\bf k}}\})\;  \label{a10}
\end{equation}
(see Appendix B). Here $\Delta _{CF}^{(0)}$ is the energy of the lowest CF
excitation which is introduced for the sake of convenience to make all
matrices dimensionless), $\{n_{{\bf k\nu }}\}$ is the set of average
occupation numbers $n_{{\bf k}\nu }$ which obeys the mean-field global
constraint condition 
\begin{equation}
N^{-1}\sum_{{\bf k}\nu }n_{{\bf k}\nu }=1\quad .  \label{a101}
\end{equation}
The occupation numbers 
\begin{equation}
n_{{\bf k}\nu }=\left[ 1+\exp \left\{ \frac{\Delta _{CF}^{(0)}\Phi _{{\bf k}%
\nu }-\mu }{k_{B}T}\right\} \right] ^{-1}  \label{a99}
\end{equation}
are defined in terms of form factors $\Phi _{{\bf k}\nu }$ 
\begin{equation}
\Phi _{{\bf k}\nu }=\sum_{\xi \xi ^{\prime }}\sum_{\Lambda \Lambda ^{\prime
}}\Theta _{\nu }^{\Lambda }(\xi ,{\bf k}){\cal Z}_{\xi \xi ^{\prime
}}^{\Lambda \Lambda ^{\prime }}({\bf k})\left[ \Theta _{\nu }^{\Lambda
^{\prime }}(\xi ^{\prime },{\bf k})\right] ^{*}\;.  \label{a11}
\end{equation}
( $\mu $ is the chemical potential). Then the matrix ${\sf Z}$ represented
by its matrix elements 
\begin{equation}
{\cal Z}_{\xi \xi ^{\prime }}^{\Lambda \Lambda ^{\prime }}({\bf k}%
)=F^{\Lambda \Lambda ^{\prime }}\delta _{\xi \xi ^{\prime }}+\frac{1}{2}%
\sum_{u}e^{i{\bf ku}}\Delta _{\xi \xi ^{\prime }}^{\Lambda \Lambda ^{\prime
}}({\bf u}),  \label{a12}
\end{equation}
should be diagonalized to find the eigenvectors $\Theta _{\nu }^{\Lambda
}(\xi ,{\bf k})$. Here ${\bf u}={\bf l}-{\bf l}^{\prime }$. The matrix $%
F^{\Lambda \Lambda ^{\prime }}$ has the form 
\begin{equation}
F^{\Lambda \Lambda ^{\prime }}=\delta _{\Lambda \Lambda ^{\prime }}\left(
E_{\Lambda }+{\cal B}^{\Lambda \Lambda }+\sum_{{\bf l}^{\prime }\xi ^{\prime
}}^{\prime }I_{\xi \xi ^{\prime }}^{\Lambda }({\bf l}-{\bf l}^{\prime
})\Delta _{CF}^{(0)}\right) /\Delta _{CF}^{(0)}+{\cal B}^{\Lambda \Lambda
^{\prime }}/\Delta _{CF}^{(0)}  \label{aaaa5}
\end{equation}
$(I_{\xi \xi ^{\prime }}^{\Lambda }({\bf l}-{\bf l}^{\prime })$ is the
dimensionless exchange integral, see Appendix A.) Finally, the variables $%
\Delta _{\xi \xi ^{\prime }}^{\Lambda \Lambda ^{\prime }}({\bf u})$
describing the RVB state [cf. eq. (\ref{11.3})] are to be obtained
self-consistently from the system of equations (\ref{a13}).

Thus, to calculate the thermodynamic coefficients of CeNiSn we use the
following procedure.

(i) We find the eigenstates $\varepsilon_{{\bf k\nu}}$ of the matrix $\Phi_{%
{\bf k}\nu}$ which depend on the parameters of the Hamiltonian $H^s$.

(ii) These eigenstates are used to calculate the imaginary part of the
correlation function ${\cal K}({\bf Q},\omega)= \langle {\bf J}_{{\bf Q}}%
{\bf J}_{{\bf -Q}} \rangle_{\omega} $ which determines the dynamic magnetic
response of the system, 
\begin{equation}
{\rm Im}{\cal K}^{\alpha\beta}({\bf Q},\omega)=\sum_{\nu\nu^{\prime}}\sum_{%
{\bf k}} n_{{\bf k\nu}}(1- n_{{\bf k-Q,\nu}}) \left\langle {\bf k}\nu \left| 
\hat{{\bf J}}_{\alpha}^+ \right| {\bf k+Q},\nu^{\prime}\right\rangle
\left\langle {\bf k+Q},\nu^{\prime}\left| \hat{{\bf J}}_{\beta} \right| {\bf %
k}\nu\right\rangle \delta \left( \hbar \omega + E_{{\bf k\nu}} - E_{{\bf k+Q}%
,\nu^{\prime}} \right)  \label{11.6}
\end{equation}
and, therefore, scattering function of magnetic neutron scattering \cite
{KAKM97}.

(iii) The average energy ${\cal E}$ is determined by the Fourier component
of this correlation function taken at zero moment 
\begin{equation}
{\cal K}^{(0)}_{{\bf Q}}=\int d{\bf k}d\omega {\cal K}^{\alpha\beta}({\bf k,Q%
},\omega)  \label{11.7}
\end{equation}
(see eq. {\ref{11.4} and Appendix B). Being diagonalized in terms of the
eigenstates $|{\bf k}\nu\rangle$, this energy is given by eq. (\ref{a10}). }

(iv) Then, using experimental results of neutron scattering and
thermodynamic measurements we fit the model parameters to describe the
neutron scattering function in absolute units and the heat capacity. Since
the heat capacity, unlike the neutron scattering spectra, is sample
dependent we obtained two set of parameters. The first set describes the
data for high quality samples and the second one corresponds to the sample
used in dilatometric measurements.

The uniform static susceptibility $\chi (T)$ characterizes the thermodynamic
response to an external magnetic field and its volume dependence
(magnetostriction is determined by the limiting value of ${\cal K}(0,0)$. We
calculate it from the relation $\chi (T)=M/B$ where $B$ is magnetic field,
and $M$ is the magnetization of the spin liquid. Only the Zeeman mechanism
of this magnetization is taken into account in the case of weak fields ($\mu
_{B}B$ less than the characteristic coupling parameters which determine the
spin-fermion spectrum).

\section{Thermodynamic relations}

We suppose that the low-temperature thermodynamics of CeNiSn is determined
mainly by the spin-liquid excitations. Since the spin liquid is an
unconventional Fermi liquid, and since our treatment of the spin-liquid
state inherits some of the shortcomings of the mean-field approximation, we
start the discussion of the thermodynamic relations with a more detailed
analysis of spin entropy. It is well known \cite{Abrik65} that one should
take special precautions to eliminate the unphysical states when introducing
the fermionic representation for the spin operators (e.g., the states doubly
occupied by fermions with opposite spin projections which are absent in
original spin representation should be excluded). Without such exclusion the
wrong temperature behavior of entropy $S(T)$ will result in incorrect
description of the specific heat and other coefficients which are connected
with the specific heat by strict thermodynamical relations.

To verify the applicability of our approach we compared the number of states
in our model with that in the usual Fermi liquid. The model situations
considered in Appendix B demonstrate non-universality of the $S(T)$ law and
its crucial dependence on the parameters of the Hamiltonian, and, in
particular, on the character of admixing the magnetic CF excitations to the
lowest Kramers doublet in the course of forming the spin-fermion branch of
the excitation spectrum. The diagonalization procedure described above gives
the equation for the average energy which is sensitive both to the
degeneracies of the bare states of the Hamiltonian $H_{f}$ which are lifted
by the spin-liquid correlations, and to the temperature compared with the
degeneracies lifted already in $H_{f}$ (CF level splitting). According to
the calculations for the third model of Appendix B which is close enough to
the real situation in CeNiSn (see below) the average energy of the spin
liquid (\ref{a10}) in the Kondo lattice with several sublattices and several
Kramers doublets involved can be presented in the form 
\begin{equation}
{\cal E}(T)=\frac{\Delta _{CF}^{(0)}}{NL}\frac{1}{2}\kappa (T)\sum_{{\bf k}%
\nu }n_{{\bf k}\nu }\varepsilon _{{\bf k}\nu }(\{n_{{\bf k}}\})\;
\label{2.1}
\end{equation}
where $\Delta _{CF}^{0}=E_{E}-E_{G}$, and the factor $\frac{1}{2}\kappa (T)$
reflects the abovementioned fundamental difference between the spin-fermion
state and the conventional Fermi liquid.

This factor is essentially nonuniversal: it depends both on the degeneracy
of the low-energy branches and the lattice geometry. We simplify our
consideration by adopting a single value of this parameter for a given
geometry of the lattice and given set of model Hamiltonian parameters. As is
shown in Appendix B, the main quantity which predetermines the effective
value of $\kappa $ at low temperatures, $k_bT\ll \Delta _{CF}^{(0)}$, is the
degeneracy lifted by spin liquid correlations. According to the results of
description of the neutron scattering spectra in CeNiSn \cite{KAKM97}, the
''hidden'' degeneracy of the spectrum equals 4 because (i) all Ce ions are
in equivalent crystallographic positions, (ii) the mixing interactions
comparable in magnitude with $\Delta _{CF}^{(0)}$ connects only Ce ions
belonging to the same $bc$-planes (see fig. 4) although the CeNiSn lattice
formally has four sublattices, (iii) only one excited state $|E\pm \rangle
=|\pm 3/2\rangle $ interplays with the ground state $|G\pm \rangle =a|\pm
1/2\rangle \pm b|\mp 5/2\rangle $ which is responsible for the formation of
the spin-fermion branch of the spectrum. As a result we come to the
situation with two Kramers doublets and two sublattices which is treated in
details in the last example of Appendix B. Therefore we adopt the value of $%
\kappa =1/4$ at $k_bT\ll \Delta _{CF}^{(0)}$.

At high temperatures the admixing of higher states (the magnetic CF excitons
and the branches split due to inter-sublattice exchange) becomes essential.
As a result, the estimations of the coefficients in equation (\ref{b10})
together with the normalization condition (\ref{a81}) give the value of $%
\kappa \approx 1/2$ for these temperatures. Eventually, at high enough
temperatures exceeding all energy splittings in our Kondo lattice the normal
behavior ${\cal S}(T)$ is restored, but the apparent \cite{Tak90} entropy
deficite at low and intermediate temperatures is an intrinsic property of
the model. This deficite is an observable effect and its existence was
noticed in many measurements (see below).

The spinon contribution to the heat capacity $c_{V}(T)$ per mole of Ce ion
for fixed volume $V$ was calculated using the standard expression 
\begin{equation}
c_{V}(T)=N_{A}(\partial {\cal E}(T)/\partial {T})_{V}  \label{aa1}
\end{equation}
where ${\cal E}(T)$ is the spinon energy per magnetic ion (\ref{2.1}) and $%
N_{A}$ is the Avogadro number. The molar spinon entropy for fixed volume $%
{\cal S}_{m}$ and the free energy of the mole ${\cal F}_{m}(T)$ are found
from equations 
\begin{eqnarray}
{\cal S}_{m}(T) &=&\int_{0}^{T}\frac{c_{V}(\tau )}{\tau }d\tau \ ,
\label{aa2} \\
{\cal F}_{m}(T) &=&N_{A}{\cal E}(T)-T{\cal S}_{m}(T)\ .
\end{eqnarray}

It is known that, in general, such thermodynamic characteristics as the
magnetic susceptibility, the volume expansion coefficient and the volume
magnetostriction can be estimated from the dependence of free energy ${\cal F%
}_{m}(T,V,B)$ on the volume $V$ and magnetic field $B$. We find these
dependences within a framework of the model of spinon spectrum which was
successfully used in \cite{KKM96} for the description of inelastic neutron
scattering spectrum.

\section{Model of spinon spectrum}

CeNiSn crystallizes in the orthorhombic lattice which belongs to the
noncentrosymmetric space group $Pn2_{1}a$ \cite{Higash93}. The point
symmetry of the crystal field on Ce ions can be treated as nearly trigonal ($%
D_{3d}$) with the rotation axis parallel to the a-axis of the crystal, and
the monoclinic distortion ($C_{s}$) can be considered as a small correction
to the trigonal crystal field \cite{Aleks94}. Therefore, to describe the
bare CF states $\Lambda $ we use the irreducible representation of the
trigonal point group $D_{3d}$. It is shown by indirect experiments \cite
{Aleks94} and confirmed by the quantitative agreement of the calculated and
experimental inelastic neutron scattering spectra \cite{KAKM97} that the
ground state level and the first excited level form a pair of Kramers
doublets 
\begin{eqnarray}
\left| G\pm \right\rangle &=&a\left| \pm 1/2\right\rangle \pm b\left| \mp
5/2\right\rangle \ ,  \label{2.1a} \\
\left| E\pm \right\rangle &=&\left| \pm 3/2\right\rangle  \label{2.1b}
\end{eqnarray}
separated by the energy interval $\Delta _{CF}<40K$ which is much less than
the energy of the second excited CF level of Ce $\Delta _{CF}^{(2)}$. The 
recently reported excitations centered around 40 meV \cite{Mcew98}, 
apparently should be ascribed to this second CF state.

To calculate the average energy ${\cal E}(T)$ one should solve the system of
equations (\ref{a11},\ref{a12},\ref{aaaa5},\ref{a13}) under the constraint (%
\ref{a101}). We are interested in the low-T thermodynamics of CeNiSn, ($T<20$%
K). These temperatures are essentially less than the bandwidth of the RVB
band ($W\approx 150$K, see \cite{KAKM97}), so we can treat approximately the
anomalous averages $\Delta ^{GG}$ as temperature independent correlators 
\cite{TaKuFu91,GreLa92}.

It was shown in \cite{KAKM97} that the interactions which form the
dispersion of the spin liquid excitations are confined mainly within the $bc$
plane of the CeNiSn lattice. The structure of this plane is determined by
the orthorhombic 2D elementary cells which contain two Ce ions in the sites $%
{\bf i}={\bf l\xi }$ where $\xi =1,2$ is the sublattice index (see fig. 4).
This network is defined by the Bravais vectors ${\bf B}=(b,0)$ and ${\bf C}%
=(0,c)$ and the basis vector ${\bf d}=(0,-b/2,c/2-{\cal O})$. Here ${\cal O}$
is the orthorhombic distortion which transforms the one-ion hexagonal
lattice into the two-ion orthorhombic one.

To describe the 2D spinon spectrum {\it at low temperatures} $T\to 0$ one
has to introduce the coupling constants which describe the matrix ${\cal Z}%
_{\xi \xi ^{\prime }}^{\Lambda \Lambda ^{\prime }}({\bf k})$ (\ref{a12}). We
confine ourselves with the simplest nearest neighbours (nn) approximation
and introduce the parameters ${\cal T}_{{\bf ii^{\prime }}}$ which describe
the in-sublattice 
\begin{equation}
{\cal T}_{1}=\Delta _{CF}^{(0)}\Delta _{11}^{GG}(nn)/2=\Delta
_{CF}^{(0)}\Delta _{22}^{GG}(nn)/2  \label{2.201}
\end{equation}
and inter-sublattice 
\begin{equation}
{\cal T}_{2}=\Delta _{CF}^{(0)}\Delta _{12}^{GG}(nn)/2=\Delta
_{CF}^{(0)}\Delta _{21}^{GG}(nn)/2  \label{2.202}
\end{equation}
coupling. These constants are responsible for the formation of the spinon
spectrum which arise due to RVB correlations within the ground state CF
level $\Lambda =G$ \cite{KAKM97}. Similarly, the interplay of spinons with
the lowest excited CF state $\Lambda =E$ is defined by the intra-site
nondiagonal matrix element given by eq. (\ref{aaaa5}) 
\begin{equation}
{\cal G}_{1}=F^{GE}=F^{EG}  \label{2.301}
\end{equation}
and the inter-site mixing coefficient 
\begin{equation}
{\cal G}_{2}=\Delta _{CF}^{(0)}\Delta _{\xi \xi ^{\prime
}}^{GE}(nn)/2=\Delta _{CF}^{(0)}\Delta _{\xi \xi ^{\prime
}}^{EG}(nn)/2;\;\xi \ne \xi ^{\prime }  \label{2.302}
\end{equation}
The condition $\xi \ne \xi ^{\prime }$ for intermixing of ground and excited
doublets arises due to the orthorhombic distortion which results in the
two-sublattice structure of the plane (Fig. 4).

To introduce the renormalised CF splitting $\tilde{\Delta}_{CF}$ one has to
consider the diagonal terms of eq. (\ref{aaaa5}). This contribution 
\begin{equation}
\tilde{\Delta}_{CF}=\Delta _{CF}^{(0)}\left( F^{EE}-F^{GG}\right)
\label{2.402}
\end{equation}
renormalises the bare value of CF splittng $\Delta _{CF}^{(0)}=E_{E}-E_{G}$
Then two terms determine the renormalization of CF splitting $\Delta
_{CF}^{0}$ 
\begin{equation}
\tilde{\Delta}_{CF}=\Delta _{CF}^{(0)}+\delta _{1}+\delta _{2}\ .
\label{rede}
\end{equation}
The first correction 
\begin{equation}
\delta _{1}={\cal B}^{EE}-{\cal B}^{GG},  \label{de1}
\end{equation}
is defined by intra-site processes, and the second one is due to inter-site
interaction 
\begin{equation}
\delta _{2}=\sum_{{\bf l}^{\prime }\xi ^{\prime }}^{\prime }\left\{ I_{\xi
\xi ^{\prime }}^{E}({\bf l}-{\bf l}^{\prime })-I_{\xi \xi ^{\prime }}^{G}(%
{\bf l}-{\bf l}^{\prime })\right\} \ .  \label{de2}
\end{equation}

Details of the procedure which diagonalizes the matrix ${\cal Z}%
_{\xi\xi^{\prime}}^{\Lambda\Lambda^{\prime}}({\bf k})$ and gives the
eigenvectors $\Theta_{\nu}^{\Lambda}(\xi,{\bf k})$ and eigenvalues $\Phi_{%
{\bf k}\nu}$ are described in \cite{KAKM97}.

\section{Sample dependence of thermodynamic properties}

It is known that the thermodynamic properties of CeNiSn depend on the
specimen quality \cite{Naka95}. At the same time the main features of
inelastic magnetic scattering spectra are the same (i.e. 2.5 meV and 4 meV
inelastic excitations) for different specimens \cite{Mason92,Sato95,Kambe96}%
. Therefore, the theory which consider the thermodynamics should explain
both the sample dependent thermodynamic properties and the sample
independent neutron scattering function.

First, it is known that the more imperfect is the sample of CeNiSn, the
higher is the residual resisitivity \cite{Taka96} and the lesser is the mean
free path of the electrons $\lambda $. The increase of $\lambda $ in more
perfect crystals results in the decrease of the exchange RKKY-type
interaction $I(R_{\mbox{Ce-Ce}})\sim \exp (-R_{\mbox{Ce-Ce}}/\lambda )$ \cite
{Mattis} which determines the parameters ${\cal T}_{1}$ and ${\cal T}_{2}$.
Therefore these parameters should be smaller for less perfect samples.
Moreover, since the relative change of the resisitivity is higher along the $%
c$-axis than along the $b$-axis, the suppression of the exchange interaction
along the $c$-axis should be larger. Hence, the reduction of ${\cal T}_{2}$
is greater than that of ${\cal T}_{1}$.

Second, the increase of the defect concentration results in the increase of
the inter-site mixing parameter ${\cal G}_{1}$ due to lowering of the
lattice symmetry. The change of the inter-site mixing parameter ${\cal G}%
_{2} $ is influenced both by increase of defect induced mixing and by
reduction of inter-site exchange. Therefore, the influence of imperfections
on this parameter is not known a priori. Since it is supposed \cite{KAKM97}
that $\left| I_{\xi \xi ^{\prime }}^{G}({\bf l}-{\bf l}^{\prime })\right|
\gg \left| I_{\xi \xi ^{\prime }}^{E}({\bf l}-{\bf l}^{\prime })\right| $,
the renormalisation of CF splitting $\tilde{\Delta}_{CF}-\Delta _{CF}^{(0)}$
is defined mainly by the exchange interaction $I_{\xi \xi ^{\prime }}^{G}(%
{\bf l}-{\bf l}^{\prime })$. One of the conditions of spin liquid RVB state
formation is the positive antiferromagnetic sign of these interactions.
Therefore, the renormalised value of CF $\tilde{\Delta}_{CF}$ is lower for
better samples.

The last effect which has to be considered is the change of the coefficients
of the wave functions (\ref{2.1a}). This change is connected with the
renormalization of the wave functions of low-symmetry system when the
exchange interaction changes \cite{Mis97}.

Two sets of parameters which take into account these tendencies for the high
quality specimen \cite{Taka96} and the specimen which was used in linear
expansion and magnetostriction measurements \cite{KVBT94,Nolt95} are
presented in Table 1. The fragments of DOS for both sets of parameters are
shown in Fig. 5. The heat capacity for perfect and imperfect specimens are
presented in Fig. 6. To illustrate the relative insensibility of neutron
spectra to the sample quality we calculated the scattering function (in
absolute units) for momentum transfers where 2.5 meV (Fig. 7a) and 4 meV
(Fig. 7b) are observed. It is seen that the neutron scattering spectra for
both specimens coincide in gross features. The experimental absolute values 
\cite{Sato95} of the scattering cross sections are also reproduced. More
detailed description of the influence of sample quality on the properties of
CeNiSn will be presented in a forthcoming publication.

In the following analysis we use the set of parameters presented in the
second column of Table 1 which correspond to the heat capacity of the less
perfect specimen used in dilatometric measurements \cite{KVBT94,Nolt95}.

\section{Magnetic susceptibility}

The standard definition 
\begin{equation}
\chi _{m}(T)=-\frac{1}{B}\left( \frac{\partial {\cal F}_{m}(T)}{\partial B}%
\right) _{T,N}  \label{susc}
\end{equation}
was used in the calculation of spinon contribution to molar magnetic
susceptibility $\chi _{m}.$ Since the magnetic response of CeNiSn is maximal
for field $B_{a}$ applied along $a$-direction \cite{Taka93}, we consider
only this easy axis component of $\chi _{m}$ and limit ourselves by Zeeman
mechanism of polaization of spin liquid described by the Hamiltonian 
\begin{equation}
\hat{H}_{Z}=g_{J}\nu _{B}\hat{J}_{a}B_{a}\ .  \label{zee}
\end{equation}
Here $g_{J}=6/7$ is the Lande factor for Ce$^{3+}$ configuration, $\nu _{B}$
is Bohr magneton and $\hat{J}_{a}$ is the $a$-projection of total angular
momentum operator. It is known \cite{Taka93} that the Sommerfeld coefficient 
$\gamma _{sp}\equiv C(T)/T$ is practically constant at lowest temperatures $%
T<1$K. Therefore we use the Fermi liquid relations for spinon subsystem at $%
T\to 0$. For example, the Wilson ratio for spinon liquid can be derived in
close analogy with electron Fermi liquid. expression. The spinon wave
function in the band $\nu $ with a wave vector {\bf k} can be represented as
a linear combination 
\begin{equation}
\mid \nu {\bf k}\rangle =\sum_{\lambda =1}^{6}{\cal L}_{\lambda }(\nu ,{\bf k%
})\left| \frac{2\lambda -7}{2}\right\rangle ,  \label{wf}
\end{equation}
where ${\cal L}_{\lambda }(\nu ,{\bf k})$ are coefficients which obey the
orthonormality relations. Therefore, applying the Zeeman operator (\ref{zee}%
) to this state we calculate the factor $G_{sp}$ 
\begin{equation}
G_{sp}=\left( \sum_{\lambda =1}^{6}\left| {\cal L}_{\lambda }(\nu ,{\bf k}%
)\right| ^{2}\frac{2\lambda -7}{2}\right) ^{2}
\end{equation}
which appears in the Wilson ratio instead of the electron $g$-factor: 
\begin{equation}
\chi _{m}(T\to 0)=\frac{3}{\pi ^{2}}\frac{\mu _{B}^{2}}{k_{B}^{2}}%
G_{sp}\gamma _{sp}\;.  \label{wilson}
\end{equation}

Neglecting in the simplest approximation admixture of the state $\mid E\pm
\rangle =\mid \pm 3/2\rangle $ to the spinons which form the RVB band, we
find that the lowest spinon state generated by the level (\ref{2.1a}) gives 
\begin{equation}
G_{sp}=g_{J}^{2}\left( 5b^{2}/2-a^{2}/2\right) ^{2}.  \label{gsp}
\end{equation}
Then, using the parameters from the last column of table 1, we find the
value of 1.34$\cdot 10^{-3}$ emu/mol for the spinon magnetic susceptibility
at $T\to 0$. This calculated value is significantly lower than the measured
one. Therefore, one should conclude, that either magnetic susceptibility is
determined not only by spinon contribution, or that the mechanism of
magnetic field influence on the spin liquid can not be reduced to simple
Zeeman splitting.

To resolve this alternative, we calculated the temperature dependence of
magnetic susceptibility up to $T=20$K by direct use of eq. (\ref{susc}). It
is seen (Fig. 8) that good agreement with experiment can be obtained if one
consider the total magnetic susceptibility as a sum of spinon Zeeman part $%
\chi _{sp}(T)$ and backgroung contribution $\chi _{b}=6.35\cdot 10^{-3}$
emu/mol which is constant at $T<20$K. This calculation perfectly reproduces
the position of maximum and the shape of the curve at $T<20$K. Since $\chi
_{sp}(T)$ in this temperature interval is determined by the low-energy sharp
features of spinon spectrum one can conclude that the Zeeman splitting of
the structured part of spinon spectrum is responsible for observed behavior
of $\chi (T)$ . The contribution $\chi _{b}$ is, apparently, detemined by
energy scales which are higher than the CF splitting. Possible sources of
this contributions will be discussed in concluding section 9.

It should be noted that the spinon response perfectly reproduces the
intensities of the 2.5 meV and 4 meV peaks in the inelastic magnetic
scattering spectra of neutrons in {\it absolute units}. Both peaks are
connected with the low energy structured part of spinon spectrum. This
observation gives one more evidence in favor of the assumption that the
constant contribution to static magnetic susceptibility is connected with
larger energy scales.

More evidences in favor of additional contributions to magnetic response are
provided by detailed study of thermal expansion and magnetostriction
presented in the next section.

\section{Temperature and field dependence of lattice distortion}

\subsection{Thermal expansion}

The \hspace{0pt}conventional phenomenological analysis of the thermal
expansion of Kondo lattices is based on the assumption that the
characteristic temperature $T_{K}$ scales all thermodynamic quantities at
low $T$, so the main contribution to the volume dependence of these
quantities may be characterized by a Gr\"{u}neisen parameter $\gamma
_{T_{K}}=$ $\partial \log T_{K}/\partial \log V$. We have seen that the
interplay between heavy fermions and crystal field excitations introduces an
additional characteristic energy scale $\tilde{\Delta}_{CF}$ and a
corresponding coupling constant, so this interplay rules out a possibility
of being content with a single scaling parameter. Moreover, the energy $%
T^{*} $ is, apparently, not related directly to $T_{K}$ in the CeNiSn
family. Thus we start this section with the derivation of Gr\"{u}neisen
parameters which characterize the spin liquid in Kondo lattices with soft CF
excitations, still confining ourselves to $T\leq \Delta _{CF}\ll T_{K}$.The
volume thermal expansion coefficient 
\begin{equation}
\alpha _{V}=\left( \frac{\partial \log V}{\partial T}\right) _{P}
\label{aa3}
\end{equation}
($P$ is pressure) can be expressed in terms of the isothermal
compressibility $\kappa _{T}=(\partial \log V/\partial P)_{T}$ and the
isothermal derivative of the total entropy ${\cal S}_{V}^{tot}$ with respect
to the volume $V$ \cite{Varley56}. 
\begin{equation}
\alpha _{V}=\kappa _{T}\left( \frac{\partial {\cal S}_{V}^{tot}}{\partial V}%
\right) _{T}.  \label{aa4}
\end{equation}
Since the Cornut-Coqblin transformation decouples spin and charge degrees of
freedom, the total entropy can be expressed as a sum ${\cal S}_{V}^{tot}=%
{\cal S}_{V}^{sp}+{\cal S}_{V}^{el}$ of the spinon ${\cal S}_{V}^{sp}$ and
the conduction electrons ${\cal S}_{V}^{el}$ contribution. The entropy $%
{\cal S}_{V}^{el}=V\gamma ^{el}T$ of conduction electrons is proportional to
the Sommerfeld coefficient $\gamma ^{el}$ at low $T,$ which results in a
linear-$T$ law for the thermal expansion 
\begin{equation}
\frac{\alpha _{V}^{el}}{T}=\kappa _{T}{\gamma ^{el}}\left( 1+\frac{\partial
\log \gamma ^{el}}{\partial \log V}\right)  \label{aa5}
\end{equation}
(see, e.g., \cite{Varley56}).

Due to the inequality $\tilde{\Delta}_{CF}\ll T_{K}$ the spinon component of 
$\alpha _{V}^{sp}/T$ can be decomposed into temperature dependent and
constant terms. It is convenient to express the spin entropy and its
derivative (\ref{aa4}) in molar units, 
\begin{equation}
\alpha _{V}^{sp}=\kappa _{T}\left( \frac{\partial {\cal S}_{m}}{\partial V}%
\right) _{T}.  \label{aa6}
\end{equation}
The isothermal compressibility and entropy per mole ($\kappa _{T}[mJ/mol]$
and ${\cal S}_{m}[mJ/(mol\cdot K)]$) enter this equation. Since the spinon
entropy is a function of model constants ${\cal P}_{i}=\widetilde{\Delta }%
_{CF},{\cal T}_{1},{\cal T}_{2},{\cal G}_{1},{\cal G}_{2}$, one can express
its volume derivative in terms of the corresponding Gr\"{u}neisen parameters$%
.$%
\begin{equation}
\gamma _{i}^{V}=\frac{\partial \log {\cal P}_{i}}{\partial \log V}\ .
\label{aa7}
\end{equation}
Strictly speaking, the volume dependence of $\tilde{\Delta}_{CF}$ has to be
expressed in terms of magnetoelastic Hamiltonian constants \cite{CalCal65}.
Hovever, in case of significant contribution of exchange interaction the
standard magnetoelastic Hamiltonian has to be revised \cite{Mis97}.
Therefore, we prefer to describe the magnetoelastic coupling in terms of
Gr\"{u}neisen parameters, which reflect the main features of magnetoelastic
interaction.

As a result, the spinon contribution to the volume thermal expansion
coefficient acquires the form 
\begin{equation}
\frac{\alpha _{V}^{sp}}{T}=\kappa _{T}\frac{{\cal S}_{m}}{T}\sum_{i}\gamma
_{i}^{V}\left( \frac{\partial \log {\cal S}_{m}}{\partial \log {\cal P}_{i}}%
\right) _{T}.  \label{aa8}
\end{equation}
The peculiar features of $\alpha _{V}^{sp}/T$ are determined by those terms
which demonstrate appreciable temperature dependence of the logarithmic
derivatives $\partial \log {\cal S}_{m}/\partial \log {\cal P}_{i}$. The
logarithmic derivatives $\partial \log {\cal S}_{m}/\partial \log {\cal T}%
_{1,2}$, which can be expressed in terms of conventional Gr\"{u}neisen
parameter $\gamma _{T_{K}}=\partial \log T_{K}/\partial \log V$, are
temperature independent in the considered temperature range $T\ll T_{K}$.
Therefore, these terms can be incorporated into a constant $W_{V}$ together
with the temperature independent contribution of conduction electrons 
\begin{equation}
W_{V}=\kappa _{T}{\gamma ^{el}}\left( 1+\frac{\partial \log \gamma ^{el}}{%
\partial \log V}\right) +\kappa _{T}\frac{{\cal S}_{m}}{T}\gamma
_{T_{K{}}}^{V}\left( \frac{\partial \log {\cal S}_{m}}{\partial \log T_{K}}%
\right) _{T}.
\end{equation}
Finally we come to the following expression for the volume thermal expansion
coefficient 
\begin{equation}
\frac{\alpha _{V}}{T}=W_{V}+\frac{\kappa _{T}}{T}\left\{ \gamma _{\tilde{%
\Delta}_{CF}}\left( \frac{\partial {\cal S}_{m}}{\partial \log \tilde{\Delta}%
_{CF}}\right) _{T}+\gamma _{{\cal G}_{1}}\left( \frac{\partial {\cal S}_{m}}{%
\partial \log {\cal G}_{1}}\right) _{T}+\gamma _{{\cal G}_{2}}\left( \frac{%
\partial {\cal S}_{m}}{\partial \log {\cal G}_{2}}\right) _{T}\right\} \ .
\label{aa10}
\end{equation}

Alhough one can not distinguish between electron and spinon contribution
into the constant term $W_{V}$, careful analysis of temperature dependent
contributions provides important infomation about the volume dependence of
spinons-CF coupling. It is obvious that the on-site mixing strength ${\cal G}%
_{1}$ is the parameter which is less influenced by volume change than the
crystal field splitting $\tilde{\Delta}_{CF}$ and the intersite mixing
parameter ${\cal G}_{2}$. Therefore, one can assume that $\gamma _{{\cal G}%
_{1}}\sim 0$ and then analyse the thermal expansion in terms of {\it two
Gruneisen parameters}, i.e. $\gamma _{\tilde{\Delta}_{CF}}$ and $\gamma _{%
{\cal G}_{2}}$. This means that even in the temperature range $T\le \tilde{%
\Delta}_{CF}$ there are two mechanisms of the volume dependence of the
spinon spectrum. Therefore, one can expect that any attempt to describe the
volume-dependent properties of CeNiSn by means of a single Gr\"{u}neisen
parameter will result in the temperature dependence of the latter \cite
{Holt96}.

Our two-parameter procedure gives good agreement with experimental data \cite
{Nolt95} for the set of parameters $\gamma _{\tilde{\Delta}_{CF}}=-90$ and $%
\gamma _{{\cal G}_{2}}=-60$ (see Fig.\ 9). It should be noted that the signs
of both Gr\"{u}neisen parameters are reasonable, i.e. expansion of the
lattice leads to softening of $\Delta _{CF}$ and to decrease of intersite
mixing parameter ${\cal G}_{2}$.

To analyse the anisotropy of thermal expansion one has to introduce the
axis-dependent Gr\"{u}neisen parameters 
\begin{equation}
\gamma _{\tilde{\Delta}_{CF}}^{x_{j}}=\frac{\partial \log \tilde{\Delta}_{CF}%
}{\partial \log x_{j}};\;x_{j}=a,b,c  \label{aa11}
\end{equation}
and 
\begin{equation}
\gamma _{{\cal G}_{2}}^{x_{j}}=\frac{\partial \log {\cal G}_{1}}{\partial
\log x_{j}};\;x_{j}=a,b,c  \label{aa12}
\end{equation}
with the obvious property 
\begin{equation}
\gamma _{i}^{V}=\sum_{x_{j}}^{a,b,c}\gamma _{i}^{x_{j}};\;i=\tilde{\Delta}%
_{CF},{\cal G}_{1}\ .  \label{aa13}
\end{equation}

A comparison of the calculated linear expansion coefficients with
experimental data \cite{KVBT94} is presented in Fig. 10. Since the intersite
mixing parameter ${\cal G}_{2}$ can be attributed to the deviation of Ce
sublattice symmetry from hexagonal one, one can expect that the magnitude of
intersite mixing should be proportional to orthorhombic distortion of Ce
sublattice (see, e.g., Fig.\ 1 in \cite{KAKM97}). Therefore, since the
lattice expansion along the c-axis is most sensitive to the orthorhombic
distortion, the inequality $\mid \gamma _{{\cal G}_{2}}^{c}\mid \gg \mid
\gamma _{{\cal G}_{2}}^{a,b}\mid $ should be valid. Indeed, the best fit is
obtained for $\gamma _{{\cal G}_{2}}^{c}=-60$ and $\gamma _{{\cal G}%
_{2}}^{b}=\gamma _{{\cal G}_{2}}^{a}=0$. The same reasoning leads to
conclusion about opposite signs of the influence of expansion in the $ab$
plane and along the $c$ axis on the CF parameters. Indeed, the values of
fitted Gr\"{u}neisen parameters are $\gamma _{\tilde{\Delta}_{CF}}^{a}=-60$, 
$\gamma _{\tilde{\Delta}_{CF}}^{b}=-84$ and $\gamma _{\tilde{\Delta}%
_{CF}}^{c}=54$.

\subsection{Magnetostriction}

Two-parameter Gr\"{u}neisen analysis of thermal expansion can be used as the
basis for quantitative explanation of the magnetostriction which in our
model is the quantity characterising the sensitivity of spinon-CF interplay
parameters to the volume change. Reversible volume magnetostriction is
thermodynamically equivalent to the strain dependence of the magnetic
susceptibility $\chi (B,T)$ \cite{ChanFaw71} 
\begin{equation}
\lambda _{V}^{\prime }(B,T)=\kappa _{T}H\chi (B,T)\left( \frac{\partial \log
\chi (B,T)}{\partial \log V}\right) _{T}.  \label{aa15}
\end{equation}
Like the thermal expansion, $\lambda _{V}^{\prime }$ in low magnetic fields
can be decomposed in the sum of electronic $\lambda _{V,el}^{\prime }$ and
spinon contribution $\lambda _{V,sp}^{\prime }$ 
\begin{equation}
\lambda _{V}^{\prime }=\lambda _{V,el}^{\prime }+\lambda _{V,sp}^{\prime }\ .
\label{aa16}
\end{equation}
The contribution of conducton electrons is temperature independent and
linear in magnetic field for $k_{B}T\ll \varepsilon _{F}$ and $\nu _{B}H\ll
\varepsilon _{F}$. 
\begin{equation}
\lambda _{V,el}^{\prime }=\kappa _{T}W_{V}^{el}B  \label{aa17}
\end{equation}
where 
\begin{equation}
W_{V}^{el}=\left( \frac{\partial \chi _{P}}{\partial \log V}\right) _{T}
\label{aa18}
\end{equation}
(here $\chi _{p}$ is the Pauli paramagnetic susceptibility). Therefore, to
reveal peculiar features of the differential magnetostriction it is
convenient to compare experimental and theoretical results for the doubly
differential magnetostriction coefficient 
\begin{equation}
\lambda _{V}^{\prime \prime }=\frac{\lambda _{V}^{\prime }}{B}  \label{aa19}
\end{equation}
which does not depend on temperature and field in the conventional Fermi
liquid. This quantity, is proportional to the volume derivative of the
magnetic susceptibility (provided the field and the temperature dependence
of the isothermal compressibility is neglected).

The Gr\"{u}neisen analysis of the magnetostriction is similar to that of the
linear expansion. One should take into account only Gr\"{u}neisen parameters
for which the logarithmic derivative of the magnetic susceptibility
demonstrates sharp temperature and field dependence whereas the
structureless contribution due to the standard Gr\"{u}neisen parameter $%
\gamma _{T_{K}}$ can be taken into account by the renormalization of the
normal Fermi-liquid contribution constant $W_{V}^{el}$ $\rightarrow $ $%
\tilde{W}_{V}$. However, in addition to the parameters $\gamma _{{\cal G}%
_{2}}$ and $\gamma _{\tilde{\Delta}_{CF}}$ one should take into accout the
volume dependence of the coefficient $a$ in the wave function $\mid G\pm
\rangle $ (\ref{2.1a}) 
\begin{equation}
\gamma _{a}=\frac{\partial \log a}{\partial \log V}\ .  \label{aa20}
\end{equation}

Then the final expression for the logarithmic derivative of the magnetic
suceptibility acquires the form 
\begin{equation}
\frac{\partial \chi }{\partial \log V}=\tilde{W}_{V}+\left\{ \gamma _{\tilde{%
\Delta}_{CF}}\left( \frac{\partial \chi _{m}}{\partial \log \tilde{\Delta}%
_{CF}}\right) _{T}+\gamma _{{\cal G}_{2}}\left( \frac{\partial \chi _{m}}{%
\partial \log {\cal G}_{2}}\right) _{T}+\gamma _{a}\left( \frac{\partial
\chi _{m}}{\partial \log a}\right) _{T}\right\}   \label{aa21}
\end{equation}
where $\tilde{W}_{V}$ incorporates all contributions, which are temperature
independent at $T<20$K 
\begin{equation}
\tilde{W}_{V}=W_{V}^{el}+\gamma _{T_{K}}\left( \frac{\partial \chi _{m}}{%
\partial \log T_{K}}\right)   \label{aa22}
\end{equation}
The logarithmic derivatives of the magnetic susceptibility are presented in
Fig. 11. The fit of the temperature dependence of the volume
magnetostriction with $\gamma _{\tilde{\Delta}_{CF}}$ and $\gamma _{g}$
found above, $\tilde{W}=-1590$ $\cdot $10$^{-3}$emu/mol, and $\gamma _{a}=230
$ gives a reasonable agreement with experimental data at low magnetic fields
(Fig. 12). It turns out that the last term of eq. (\ref{aa21}) dominates in
magnetostriction because the contribution of the first two terms gives a
value which is significantly smaller than the experimental data.

Like in case of the magnetic susceptibilty,there is a large constant
contribution in additon to the temperature dependent term, which may be
ascribed to the excitations with the energy scale essentially exceeding that
of the structured part of the spinon spectrum. Moreover, this background,
which is present both in magnetic susceptibility and magnetostriction, does
not appear in linear expansion in zero magnetic field.

It should be mentioned that the suggested analysis is succesful only at low
magnetic fields. Meanwhile, the field dependence of the magnetostriction 
yields evidence in favor of a non-Zeeman influence of the magnetic field on 
the spinon
spectrum. The experiments in higher fields give almost temperature
independent behavior of the logarithmic derivative of the magnetic
susceptibility (see Fig. 3) whereas the Zeeman term behaves similarly both
in low and high fields.

\section{Conclusion}

The theory of spin liquid origin of the low temperature anomalies in
thermodynamical and magnetic properties of CeNiSn and related compounds
offered in \cite{KKP93,KVBT94} is based on the assumption that the pseudogap
features of these properties should be ascribed to spin excitations rather
than to a non-metallic electron spectrum. Later on, this suspicion was
confirmed by metallic-like behavior of the resistivity in samples of good
enough quality \cite{Taka96}. Nevertheless, the spin-liquid theory met the
challenge of explaining not only the low-temperature thermodynamics but also
the fascinatingly complicated picture of inelastic neutron spectra.
Explanation of the mechanism of neutron scattering offered in \cite{KAKM97}
provided us with a set of parameters which determine the spinon spectrum.
Thus we came to the quantitative picture of a spin liquid which arises as a
result of the interplay between intersite resonating valence bonds (spinons) 
and
one-site crystal field excitations. In the present paper the experimental
data on volume-dependent thermodynamical properties of CeNiSn are collected
and the quantitative theory of spin liquid is used for interpretation of
these data.

To summarize the results of the realization of the above program, one should
conclude that the hypothesis that the nonlocal spinon pairs determine the
free energy of CeNiSn in accordance with eq. (\ref{11.2}) is confirmed by
detailed quantitative consideration. These pairs, apparently, are
responsible both for the structure of low-energy excitations which determine
the spectrum of inelastic neutron scattering, and for the low-temperature
thermodynamics (specific heat, thermal expansion). As to magnetic properties
(susceptibility, magnetostriction), the real situation is, apparently, more
complicated than the picture given by the mean-field version of spin-liquid
theory. We confined ourselves by considering the Zeeman polarization of
spinon excitations. This mechanism successfully describes the temperature
dependences $\chi(T)$ and $\lambda^{\prime}(T)$ at low enough $T$ where the
spinon excitations are still well defined. It fails, however, to give the
absolute values of these quantities. Besides, the properties of CeNiSn
change radically in strong enough magnetic field, and this change is also
beyond the applicability of our theory.

At this stage we can only speculate about the presence of another energy
scale in the excitation spectrum of the spin liquid. To explain the origin of
this high-energy scale one should remember that the {\it coherent} branch of
spinon pairs gives only a small contribution to the low-energy part of the
spectral density of spin excitations ${\cal A}_s(\varepsilon)$. In the
absence of long range order the dominant contribution to ${\cal A}%
_s(\varepsilon)$ is given by the incoherent structurless continuum of
paramagnetic spin fluctuations. The closest analogy to our case is the
structure of spin-fluctuation spectrum in Cu-O planes of high-T$_c$
materials described by the $t-J$ model at small doping limit \cite
{Kane89,Khors93}. In that model the coherent branch of Zhang-Rice
spinon-holon compound particles forms the low-energy part of the spin-polaron
continuum. This branch gives a small contribution to ${\cal A}%
_{s}(\varepsilon) $ with characteristic energy scale $J$, whereas the main
part of ${\cal A}_{s}(\varepsilon)$ is formed by the incoherent structureless
spin-polaron continuum scaled by the energy $t\gg J$. Our crude enough
mean-field approximation for the spinon spectrum ignores the incoherent
continuum. Besides, the influence of magnetic field by no means can be
reduced to simple polarization of the spin liquid. We hope that the more 
refined
description of spin liquid in Kondo lattices will reveal the real role of
both external magnetic field and intrinsic antiferromagnetic fluctuations in 
the magnetic response of these systems.

\section{Acknowledgements}

The support of NWO (grant 07-30-002) and RFBR (grant 98-02-16730) is
acknowledged.

\section{Appendix A}

To calculate the thermal energy, one should find the thermodynamic average
of the sum of Hamiltonians $H_{h}$, $H_{(c)}^{RKKY}$ and $H_{(nc)}^{RKKY}$, 
\begin{equation}
{\cal E}(T)={\cal E}_{h}(T)+{\cal E}_{c}(T)+{\cal E}_{nc}(T)  \label{a1}
\end{equation}
where 
\begin{equation}
{\cal E}_{h}(T)=\frac{\Delta _{CF}^{(0)}}{NL}\sum_{{\bf l\xi }}\sum_{\Lambda
\Lambda ^{\prime }}F^{\Lambda \Lambda ^{\prime }}\langle f_{{\bf l}\xi
\Lambda }^{\dagger }f_{{\bf l}\xi \Lambda ^{\prime }}\rangle ,  \label{a2}
\end{equation}
\begin{equation}
{\cal E}_{c}(T)=\frac{\Delta _{CF}^{(0)}}{NL}\sum_{{\bf ll}^{\prime }\xi \xi
^{\prime }}^{\prime }\sum_{\Lambda }I_{\xi \xi ^{\prime }}^{\Lambda }({\bf l}%
-{\bf l}^{\prime })\langle f_{{\bf l}\xi \Lambda }^{\dagger }f_{{\bf l}%
^{\prime }\xi ^{\prime }\Lambda }\rangle \langle f_{{\bf l}^{\prime }\xi
^{\prime }\Lambda }^{\dagger }f_{{\bf l}\xi \Lambda }\rangle \;,  \label{a3}
\end{equation}
\[
{\cal E}_{nc}(T)=\frac{\Delta _{CF}^{(0)}}{NL}\sum_{{\bf ll}^{\prime }\xi
\xi ^{\prime }}^{\prime }\sum_{\Lambda \Lambda ^{\prime }}(1-\delta
_{\Lambda \Lambda ^{\prime }})\langle f_{{\bf l}\xi \Lambda }^{\dagger }f_{%
{\bf l}^{\prime }\xi ^{\prime }\Lambda ^{\prime }}\rangle \times 
\]
\begin{equation}
\left\{ \tilde{I}_{\xi \xi ^{\prime }}^{\Lambda \Lambda ^{\prime }}({\bf l}-%
{\bf l}^{\prime })\langle f_{{\bf l^{\prime }}\xi ^{\prime }\Lambda ^{\prime
}}^{\dagger }f_{{\bf l}\xi \Lambda ^{\prime }}\rangle +\left[ \tilde{I}_{\xi
^{\prime }\xi }^{\Lambda ^{\prime }\Lambda }({\bf l}^{\prime }-{\bf l}%
)\right] ^{*}\langle f_{{\bf l^{\prime }}\xi ^{\prime }\Lambda }^{\dagger
}f_{{\bf l}\xi \Lambda }\rangle \right\} \ .  \label{a4}
\end{equation}
Here N is the number of unit cells, primes in the lattice sums mean that the
diagonal terms are omitted, 
\begin{equation}
I_{\xi \xi ^{\prime }}^{\Lambda }({\bf l}-{\bf l}^{\prime })={\cal I}_{{\bf l%
}\xi ,{\bf l}^{\prime }\xi ^{\prime }}^{\Lambda \Lambda }/\Delta
_{CF}^{(0)}\;\;\;I_{\xi \xi ^{\prime }}^{\Lambda \Lambda ^{\prime }}({\bf l}-%
{\bf l}^{\prime })=\bar{{\cal I}}_{{\bf l}\xi ,{\bf l}^{\prime }\xi ^{\prime
}}^{\Lambda \Lambda ^{\prime }\Lambda ^{\prime }\Lambda ^{\prime }}/\Delta
_{CF}^{(0)}\;,  \label{a7}
\end{equation}
and the matrix $F^{\Lambda \Lambda ^{\prime }}$ is defined by the equation (%
\ref{aaaa5}). The quantities $\Delta _{\xi \xi ^{\prime }}^{\Lambda \Lambda
^{\prime }}({\bf u})$ are described by the system of equations 
\[
\Delta _{\xi \xi ^{\prime }}^{\Lambda \Lambda ^{\prime }}({\bf u}%
)=2N^{-1}\sum_{{\bf k}\nu }n_{{\bf k}\nu }e^{-i{\bf ku}}\left\{ \delta
_{\Lambda \Lambda ^{\prime }}I_{\xi \xi ^{\prime }}^{\Lambda }({\bf u}%
)\Theta _{\nu }^{\Lambda }(\xi ^{\prime },{\bf k})\left[ \Theta _{\nu
}^{\Lambda }(\xi ,{\bf k})\right] ^{*}+\right. 
\]
\begin{equation}
\left. (1-\delta _{\Lambda \Lambda ^{\prime }})\left( \tilde{I}_{\xi \xi
^{\prime }}^{\Lambda \Lambda ^{\prime }}({\bf u})\Theta _{\nu }^{\Lambda
^{\prime }}(\xi ^{\prime },{\bf k})\left[ \Theta _{\nu }^{\Lambda ^{\prime
}}(\xi ,{\bf k})\right] ^{*}+\left[ \tilde{I}_{\xi ^{\prime }\xi }^{\Lambda
^{\prime }\Lambda }(-{\bf u})\right] ^{*}\Theta _{\nu }^{\Lambda }(\xi
^{\prime },{\bf k})\left[ \Theta _{\nu }^{\Lambda }(\xi ,{\bf k})\right]
^{*}\right) \right\} \;,  \label{a13}
\end{equation}
and this system together with equations (\ref{a11}), (\ref{a12}) forms the
closed set of equations which should be solved self-consistently.

Although in the general case of low-symmetry lattices with anisotropic exchange
constants $I^{\Lambda}_{\xi\xi^{\prime}} ({\bf u})$ and $\tilde{I}%
^{\Lambda\Lambda^{\prime}}_{\xi\xi^{\prime}} ({\bf u})$, one should
introduce several variables $\Delta_{\xi\xi^{\prime}}({\bf u})$, for the
lattices with high enough symmetry one can confine oneself with a single
parameter $\Delta^{GG}$ which characterises the intersite correlations within
the lowest crystal field level.

For example, in Bravais lattices with $P$ nearest neighbours in equivalent
positions the set of parameters should be introduced, 
\begin{equation}
\Delta ^{\Lambda \Lambda ^{\prime }}=\frac{1}{z}\sum_{{\bf u}}^{nn}\Delta
^{\Lambda \Lambda ^{\prime }}({\bf u})\ .  \label{a14}
\end{equation}
However, if only the lowest crystal field states $\mid G\pm \rangle $ are
responsible for the ''anomalous'' intersite correlations described by the
parameter $\Delta ^{GG}$ 
\begin{equation}
\Delta ^{GG}=I^{G}\frac{2}{Nz}\sum_{{\bf k}\nu }\varphi _{{\bf k}}n_{{\bf k}%
\nu }\left| \Theta _{\nu }^{G}({\bf k})\right| ^{2},  \label{a15}
\end{equation}
there is no need in independent nondiagonal variables $\Delta ^{G\Lambda }$.
All of them can be expressed via $\Delta ^{GG}$ by means of the factors $%
q_{\Lambda }=\tilde{I}^{\Lambda G}/I^{\Lambda G}<1$, 
\begin{equation}
\Delta ^{\Lambda G}=q_{\Lambda }\Delta ^{GG}.  \label{a16}
\end{equation}
Thus, we assume that the symmetry of CeNiSn lattice is high enough to
restrict ourself to a single parameter $\Delta ^{GG}$ which characterises
the intersite correlations within the lowest Kramers doublet.

\section{Appendix B}

First, we demonstrate that the diagonalized form (\ref{a10}) contains the 
correct number of states, namely 2N levels for the simple case of
one-sublattice crystal with spins 1/2 in each site described by the
Hamiltonian 
\begin{equation}
H^{s}=\frac{I}{2}\sum_{{\bf ii^{\prime }}}^{{\bf i}\ne {\bf i}^{\prime
}}\sum_{\nu \nu ^{\prime }}f_{{\bf i}\nu }^{\dagger }f_{{\bf i}\nu ^{\prime
}}f_{{\bf i^{\prime }}\nu ^{\prime }}^{\dagger }f_{{\bf i^{\prime }}\nu }
\label{b1}
\end{equation}
($\nu =\pm $ are the spin projections). In this case the average energy of
the spin liquid state ${\cal E}_{c}$ is given by the equation: 
\begin{equation}
{\cal E}_{c}=\frac{I}{2}\sum_{{\bf ij}}\langle b_{{\bf ij}}b_{{\bf ji}%
}\rangle =\frac{I}{2z}\sum_{{\bf pq}}\sum_{\nu \nu ^{\prime }}\varphi ({\bf %
p-q})n_{{\bf p\nu }}n_{{\bf q\nu ^{\prime }\ \ }}.  \label{b2}
\end{equation}
Here $z$ is the coordination number for the $nn$ sphere. Using the
mean-field definition of the parameter $\Delta =\langle \Delta _{{\bf ij}%
}\rangle $ and the spinon energy $\varepsilon _{{\bf p}}$, 
\begin{equation}
\Delta \varphi _{{\bf p}}=\sum_{{\bf q}}\varphi _{{\bf p-q}}\tanh \frac{%
\varepsilon _{{\bf q}}}{2T}\ ,\;\;\;\;\varepsilon _{{\bf p}}=I\Delta
\varepsilon _{{\bf p}}\ ,  \label{b3}
\end{equation}
and the property of $\sum_{{\bf q}}\varphi _{{\bf q}}=0$, equation (\ref{b2}%
) is reduced to 
\begin{equation}
{\cal E}=I\sum_{{\bf p}}\varepsilon _{{\bf p}}n_{{\bf p}}\ .  \label{b4}
\end{equation}
This means that our problem is thermodynamically equivalent to the problem
of spinless fermions, so the limiting value of the entropy for this system
has the correct value of $S_{\infty }=N\log 2$. On the other hand, the naive
mean-field treatment of the Hamiltonian (\ref{b1}) results in the effective
Hamiltonian 
\begin{equation}
H_{MF}=I\sum_{{\bf p}}\varepsilon _{{\bf p\nu }}n_{{\bf p\nu }}-\frac{I}{2}%
Nz|\Delta |^{2}  \label{b5}
\end{equation}
which gives a wrong value of ${\cal S}_{\infty }=N\log 4$.

The nature of this discrepancy is well known. In the spin fermion
representation for the spin 1/2, ${\bf S}_{i}=f_{i\nu }^{\dagger }\hat{\sigma%
}f_{i\nu ^{\prime }}$ (where $\hat{\sigma}$ is the Pauli matrix) the local
constraint $\sum_{\nu }f_{i\nu }^{\dagger }f_{i\nu }=1$ forbids simultaneous
creation of both up and down spin fermions. Since this local constraint is
changed for the global constraint $N^{-1}\sum_{{\bf k\nu }}f_{{\bf k\nu }%
}^{\dagger }f_{{\bf k\nu }}=1$, one should find a procedure which prevents
simultaneous creation of ''particle'' and ''hole'' in the spinon spectrum
when calculating the thermodynamic functions to reproduce the correct
temperature behavior of entropy.

More complicated is the situation with the next model example of the Bravais
lattice with two CF Kramers doublets $|G\nu\rangle, |E\nu\rangle$ and two
intersite exchange coupling constants $I^{GG}$ and $I^{GE}$. The system of
equations (\ref{a13}) now describes two parameters, $\Delta^{GG}({\bf u})$
given by eq. (\ref{a9}) and 
\[
\Delta^{GE}({\bf u})=I^{GE}\frac{2}{Nz} \sum_{{\bf k}\nu} \varphi_{{\bf k}}
n_{{\bf k}\nu} \left| \Theta_{\nu}^{G}({\bf k}) \right|^2=q\Delta^{GG} 
\]
(see eq. \ref{a16}). Then the matrix ${\sf Z}$ (\ref{a12}) has the form 
\begin{eqnarray}
\frac{1}{2}\left( 
\begin{array}{cccc}
\Delta^{GG}\varphi_{{\bf k}} & q\Delta^{GG}\varphi_{{\bf k}} & 0 & 0 \\ 
q\Delta^{GG}\varphi_{{\bf k}} & 1 & 0 & 0 \\ 
0 & 0 & \Delta^{GG}\varphi_{{\bf k}} & q\Delta^{GG}\varphi_{{\bf k}} \\ 
0 & 0 & q\Delta^{GG}\varphi_{{\bf k}} & 1
\end{array}
\right)  \label{b6}
\end{eqnarray}
with normalization condition (\ref{a81}).

This case can be treated in the same way as the previous one, provided the
intermixing of ground and excited states is not too strong, i.e., when 
\begin{equation}
\frac{q^{2}}{|1-\frac{1}{2}\Delta ^{GG}\varphi _{{\bf k}}|}\ll 1\ .
\label{b66}
\end{equation}
Then the contribution ${\cal E}_{\varsigma }$ of the half-filled lowest
spin-fermion band to the energy ${\cal E}$ is 
\begin{equation}
{\cal E}_{\varsigma }=\frac{\Delta _{CF}^{(0)}}{2}N^{-1}\sum_{{\bf k}%
}\sum_{\nu =\pm }n_{{\bf k}\nu }^{\varsigma }\varepsilon _{{\bf k}\nu
}^{\varsigma }  \label{b7}
\end{equation}
where 
\begin{equation}
\varepsilon _{{\bf k}\varsigma }=\frac{1}{2}\Delta ^{GG}\varphi _{{\bf k}%
}\left( 1-\frac{q^{2}}{|1-\frac{1}{2}\Delta ^{GG}\varphi _{{\bf k}}|}\right)
\ .  \label{b8}
\end{equation}
In this case we also have the compensation of Kramers degeneracy. It should
be emphasized, however, that the second branch of the excitations generated
by the matrix {\sf Z} (\ref{b6}) is, in fact, the usual magnetic CF exciton
band modified by the interaction with the spin liquid branch, and its
contribution to the entropy can be treated in conventional manner, at least
at $k_{B}T\ll \Delta _{CF}^{(0)}$.

These two examples demonstrate that there is no universal recipe for
calculating the entropy in the systems with strong interplay between the
non-local spin-liquid excitations and the one-site CF excitations. The third
instructive example demonstrates the importance of accurate treatment of all
degeneracies which could be lifted by the spin-liquid correlations. Here we
consider the two-sublattice crystal with the crystal field resulting in two
equivalent Kramers doublets for Ce ion in each sublattice. Then the exchange
interactions in $H_{c}$ and $H_{nc}$ terms of the Hamiltonian (\ref{1.900})
are described by four parameters $I_{\xi \xi ^{\prime }}^{G}$ and $I_{\xi
\xi ^{\prime }}^{GE}$, where $\xi \xi ^{\prime }=1,2$. Then the matrix ${\sf %
Z}$ acquires the form 
\[
\frac{1}{2}\left( 
\begin{array}{cccc}
\Delta _{11}^{GG}\varphi _{{\bf k}} & \Delta _{12}^{GG}\varphi _{{\bf k}%
}^{\prime } & q\Delta _{11}^{GG}\varphi _{{\bf k}} & q^{\prime }\Delta
_{12}^{GG}\varphi _{{\bf k}}^{\prime } \\ 
\Delta _{12}^{GG}\varphi _{{\bf k}}^{\prime } & \Delta _{11}^{GG}\varphi _{%
{\bf k}} & q^{\prime }\Delta _{12}^{GG}\varphi _{{\bf k}}^{\prime } & 
q\Delta _{11}^{GG}\varphi _{{\bf k}} \\ 
q\Delta _{11}^{GG}\varphi _{{\bf k}} & q^{\prime }\Delta _{12}^{GG}\varphi _{%
{\bf k}}^{\prime } & 2 & 0 \\ 
q^{\prime }\Delta _{12}^{GG}\varphi _{{\bf k}}^{\prime } & q\Delta
_{11}^{GG}\varphi _{{\bf k}} & 0 & 2
\end{array}
\right) 
\]
(this matrix represents one of two Kramers subspaces in the block-diagonal
matrix ${\sf Z}={\sf Z}^{+}\otimes {\sf Z}^{-}$). We assume that the
in-sublattice structure factor $\varphi _{{\bf k}}$ and the in-sublattice
coupling constants $I_{11}$ are the same for both sublattices. Here relation
between non-diagonal and diagonal elements of in-sublattice and
inter-sublattice coupling constants are given by $q=I_{11}^{GE}/I_{11}^{G}$
and $q^{\prime }=I_{12}^{GE}/I_{12}^{G}$, respectively. In general case the
inter-sublattice structure factor $\varphi _{{\bf k}}^{\prime }\neq \varphi
_{{\bf k}}$. Under these assumptions $\Delta _{11}=\Delta _{22}$ and $\Delta
_{12}=\Delta _{21}$, and two independent spinon parameters are determined by
the equations 
\[
\Delta _{11}^{GG}=\frac{2}{Nz}\sum_{{\bf k\nu }}\varphi _{{\bf k}}n_{{\bf %
k\nu }}I_{11}^{G}\left| \Theta _{\nu }^{G}({1,{\bf k}})\right| ^{2}, 
\]
\begin{equation}
\Delta _{12}^{GG}=\frac{2}{Nz}\sum_{{\bf k\nu }}\varphi _{{\bf k}}^{\prime
}n_{{\bf k\nu }}I_{12}^{G}\Theta _{\nu }^{G}({2,{\bf k}})\left[ \Theta _{\nu
}^{G}({1,{\bf k}})\right] ^{*}.  \label{b10}
\end{equation}
In the case of strong intermixing ($I_{\xi \xi ^{\prime }}\approx 1$ and $%
q,q^{\prime }\approx 1$) all the coefficients $|\Theta _{\nu }^{G}(\xi ,{\bf %
k})|$ are of the same order, and each of them can be estimated as $\theta
\approx 1/2\sqrt{2}$. As a result, the contribution of the lowest branches $%
\varepsilon _{{\bf k}\nu }^{\varsigma }$ at low temperatures $k_{B}T\ll
\Delta _{CF}^{(0)}$ can be approximately represented as 
\begin{equation}
{\cal E}_{\varsigma }\approx \frac{1}{8}\Delta _{CF}^{(0)}N^{-1}\sum_{{\bf k}%
}\sum_{\nu =\pm }n_{{\bf k}\nu }^{\varsigma }\varepsilon _{{\bf k}\nu
}^{\varsigma }  \label{b11}
\end{equation}
because in this case all terms in the matrix ${\sf Z}$ related to lowest
branches $\varsigma $ contain the factors $\Delta ^{G}$, as was explicitely
demonstrated in the above example (see eq. \ref{b8}).

Again we see that the diagonalization procedure (\ref{a10}) results in a 
non-universal form of the average energy for spin liquid excitations in
comparison with corresponding equations for the conventional Fermi liquids,
at least at low temperatures.

\newpage

\begin{table}[tbp]
\begin{tabular}{|c|c|c|}
\hline
Parameters & High quality & S.Nolten et.al. \\ \hline
${\cal T}_1$ & 16.15 K & 14.40 K \\ 
${\cal T}_2$ & 24.65 K & 12.40 K \\ 
${\cal G}_1$ & 1.83 K & 3.00 K \\ 
${\cal G}_2$ & -6.60 K & -5.41 K \\ 
$\tilde{\Delta}_{CF}$ & -2.0 K & 8.0 K \\ 
a & 0.53 & 0.62 \\ \hline
\end{tabular}
\end{table}
Table 1. Parameters which define the spinon spectrum in the samples of
higher and lower quality

\newpage

\begin{center}
{\large Figure captions}
\end{center}
\noindent
Fig.1 Magnetostriction $\lambda $ of single-crystalline CeNiSn for a field
directed along the $a$-axis (bold lines) and elongation (or contraction)
along the $a$, $b$ and $c$-axis (thin solid, dotted and dashed lines,
respectively) at temperatures of 0.5 K (a), 1.4 K (b), and 4.3 K (c). \mbox{}
\newline
\mbox{} \newline
Fig.2 Coefficient of magnetostriction $\lambda ^{\prime }$ of
single-crystalline CeNiSn for a field directed along the $a$-axis. All
notations are the same as in fig. 1 \mbox{} \newline
\mbox{} \newline
Fig.3: Field and temperature dependence of the logariphmic volume derivative
of magnetic susceptibility evaluated from eq.\ (\ref{anne3}) (the isothermal
compressibility is $\kappa _{T}=1.8*10^{-11}$ m$^{2}$/N according to \cite
{Uwat94}): (a) field dependence for $T=0.5$ K (squares), $T=1.4$ K
(diamonds) and $T=4.3$ K (triangles); (b) temperature dependence for
different magnetic fields. \mbox{} \newline
\mbox{} \newline
Fig.4: Orthorhombic CeNiSn lattice, structure of $bc$-plane. Two Ce
sublattices are denoted by black and grey circles, respectively. The
orthorhombic distortion ${\cal O}$ (solid arrows) transforms the simple
hexagonal lattice into two-sublattice orthorhombic one. The in-sublattice
interaction ${\cal T}_{1}$ is denoted by solid double arrow. The
inter-sublattice interaction ${\cal T}_{2}$ and ${\cal G}_{2}$ is denoted by
dashed double arrow. \mbox{} \newline
\mbox{} \newline
Fig.5: Fragments of the density of states (DOS) of spinon spectra for high
quality sample (bold line) and imperfect sample (thin line). \mbox{} \newline
\mbox{} \newline
Fig.6: Temperature dependence of heat capacity. Dashed line represents the
spin-fermion contribution $\gamma _{sp},$ for imperfect sample, and
solid line gives the Sommerfeld coefficient with additional contribution
from conduction electrons $\gamma _{cond}=8$ mJ/mol K$^{2}$. The spinon part
calculated for high quality sample is presented in the insert. Experimental
points for perfect sample are taken from \cite{Taka96} (triangles) and those
for imperfect sample are taken from \cite{Nolt95} (squares). \mbox{} \newline
\mbox{} \newline
Fig.7: Scattering functions of inelastic magnetic neutron scattering in
absolute units calculated for high quality (bold line) and less perfect
sample (thin line). \mbox{} \newline
\mbox{} \newline
Fig.8: Calculated magnetic susceptibility (line) compared with experimental
data for specimen \# 4 from Ref. \cite{Taka96}. \mbox{} \newline
\mbox{} \newline
Fig.9: Calculated linear coefficient of volume expansion (line) compared
with experimental data \cite{Nolt95} (triangles). Insert: temperature
dependence of logarithmic derivatives of entropy. \mbox{} \newline
\mbox{} \newline
Fig.10: Calculated coefficients of linear expansion (lines) compared with
experimental data \cite{KVBT94} (points) along a- (solid line, squares), b-
(dashed line, triangles) and c-axis (dotted line, dimonds). \mbox{} \newline
\mbox{} \newline
Fig.11: Components of volume derivative of magnetic susceptibility
calculated from eq. (\ref{aa21}) \mbox{} \newline
\mbox{} \newline
Fig.12: Logarithmic derivative of magnetic susceptibilty calculated from eq.
(\ref{aa21}) (solid line) and extracted from experiment \cite{Uwat94}
(circles) .

\begin{figure}[e]
\epsfxsize=12cm
\epsfysize=17cm
\epsfbox{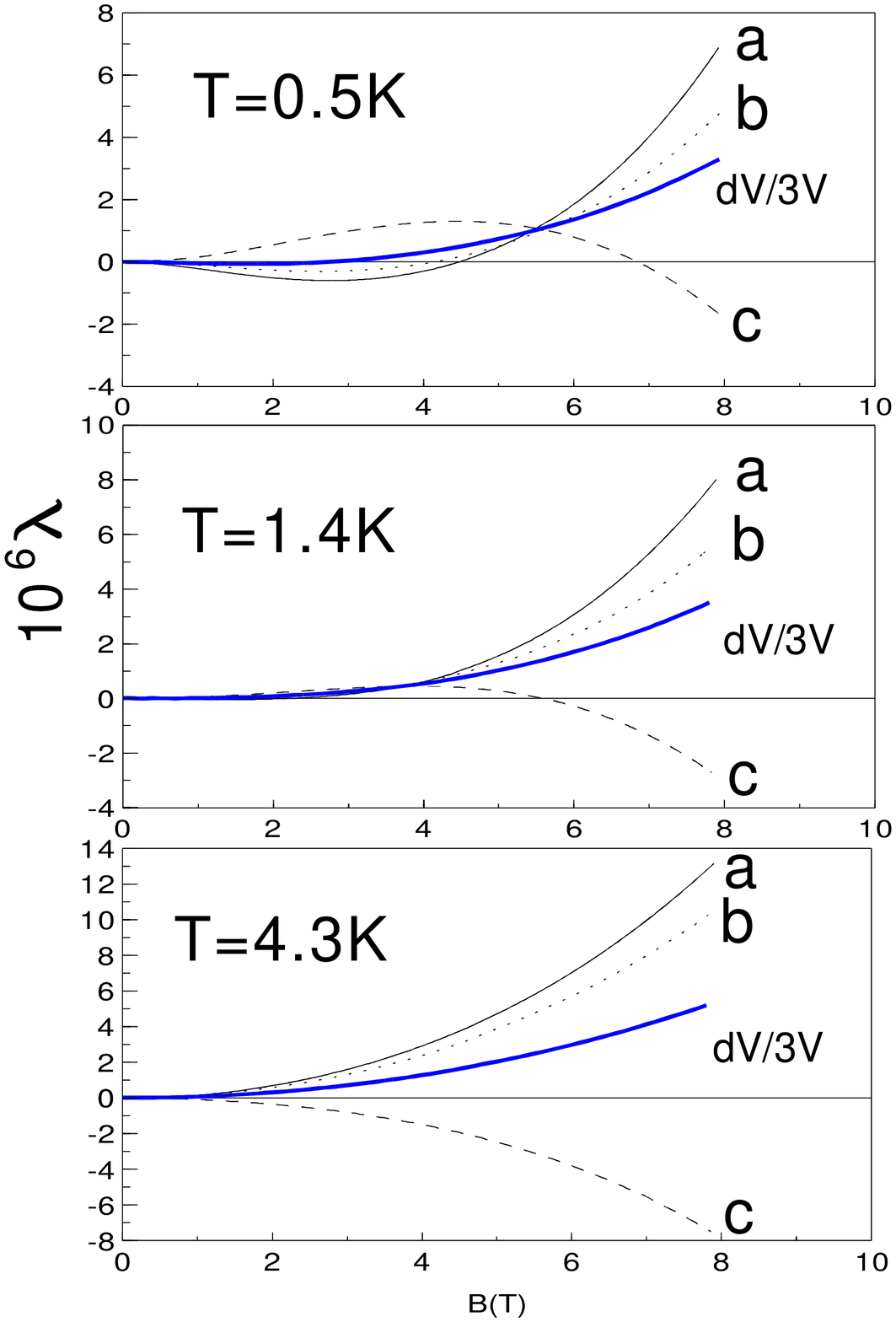}
\leavevmode
\mbox{} \\ \mbox{} \\
Fig.\ 1.  
\end{figure}

\pagebreak

\begin{figure}[e]
\epsfxsize=12cm
\epsfysize=17cm
\epsfbox{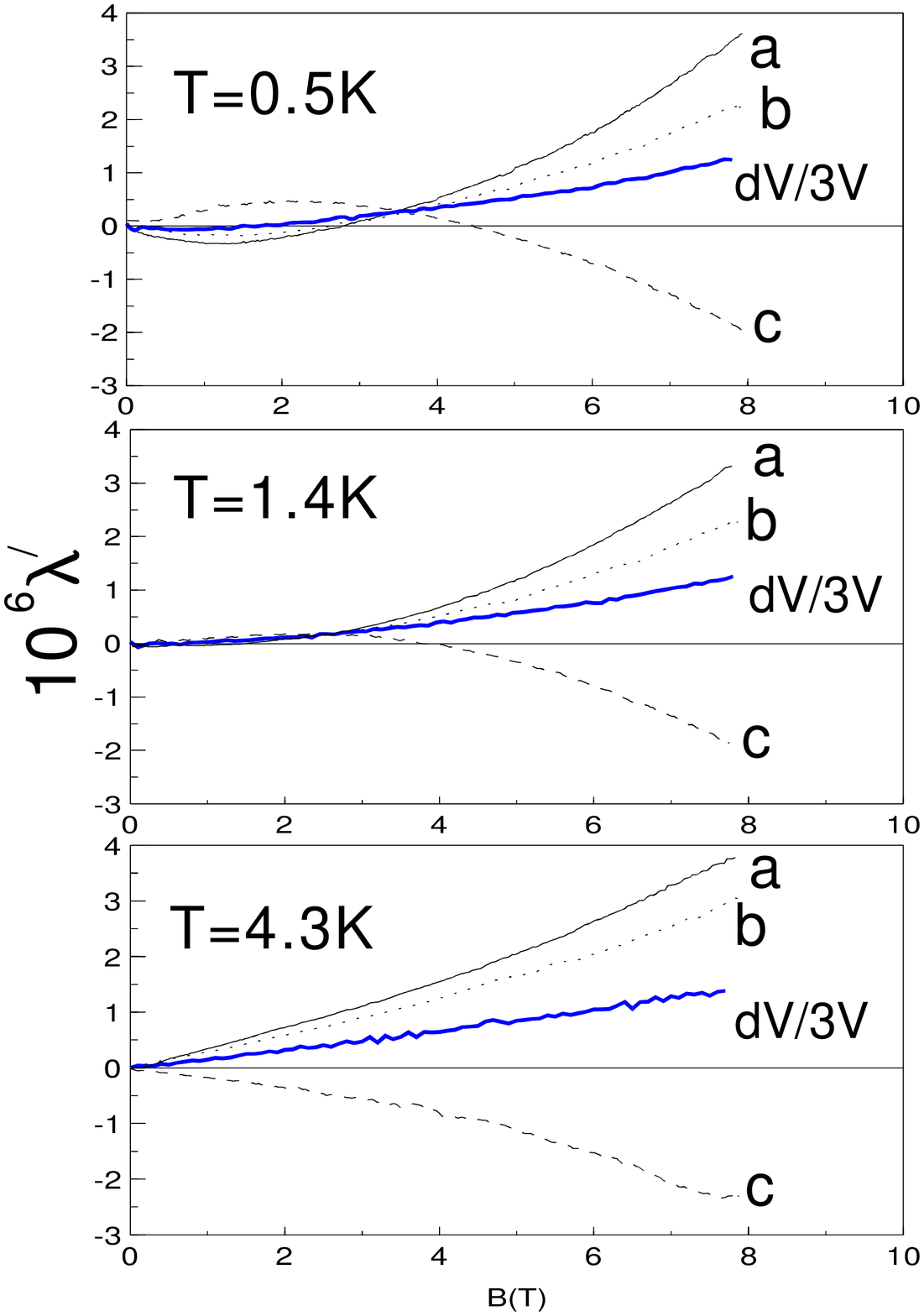}
\leavevmode
\mbox{} \\ \mbox{} \\
Fig.\ 2.
\end{figure}

\pagebreak

\begin{figure}[e]
\epsfxsize=12cm
\epsfysize=17cm
\epsfbox{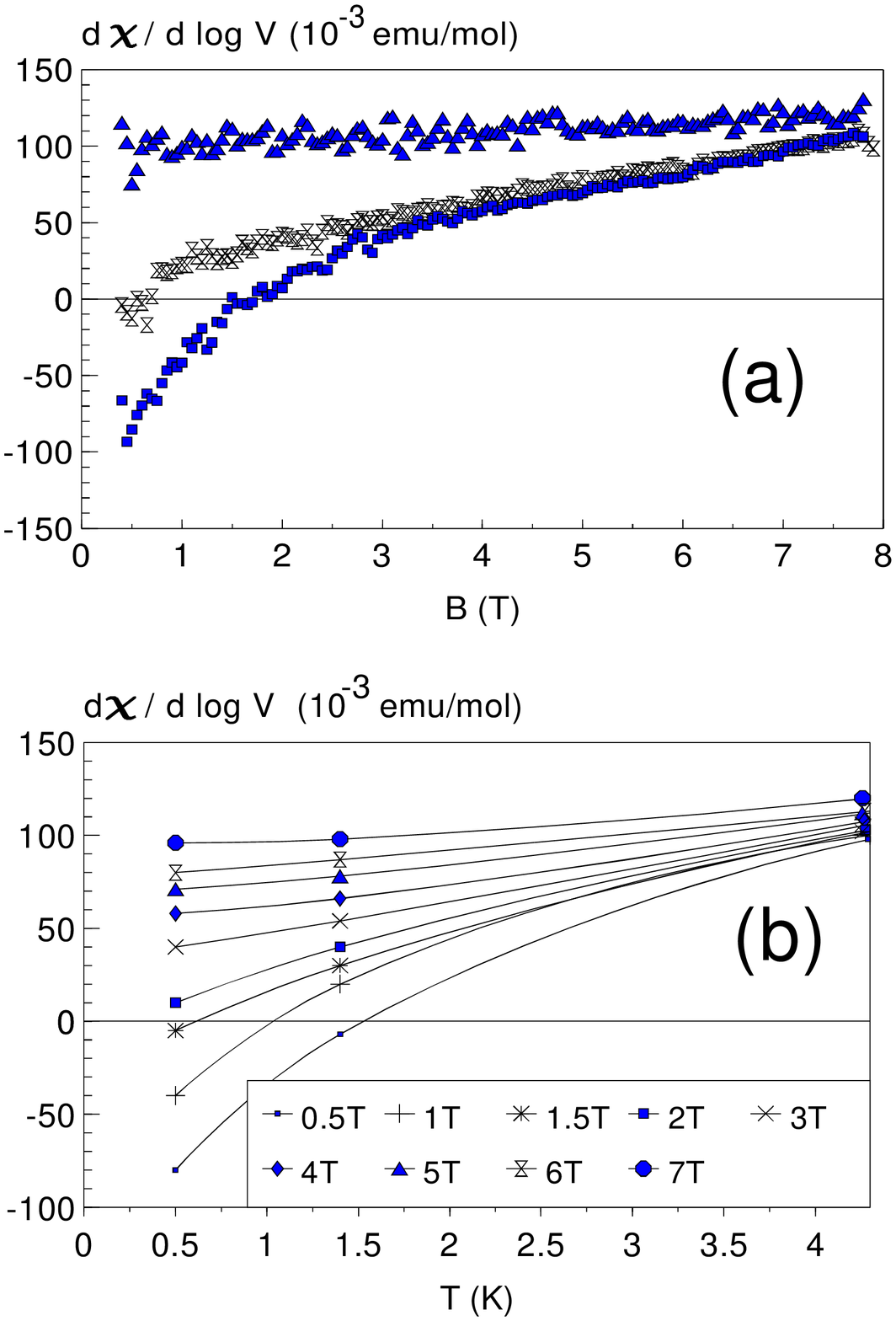}
\leavevmode
\mbox{} \\ \mbox{} \\
Fig.\ 3.
\end{figure}

\pagebreak

\begin{figure}[e]
\epsfxsize=16cm
\epsfysize=12cm
\epsfbox{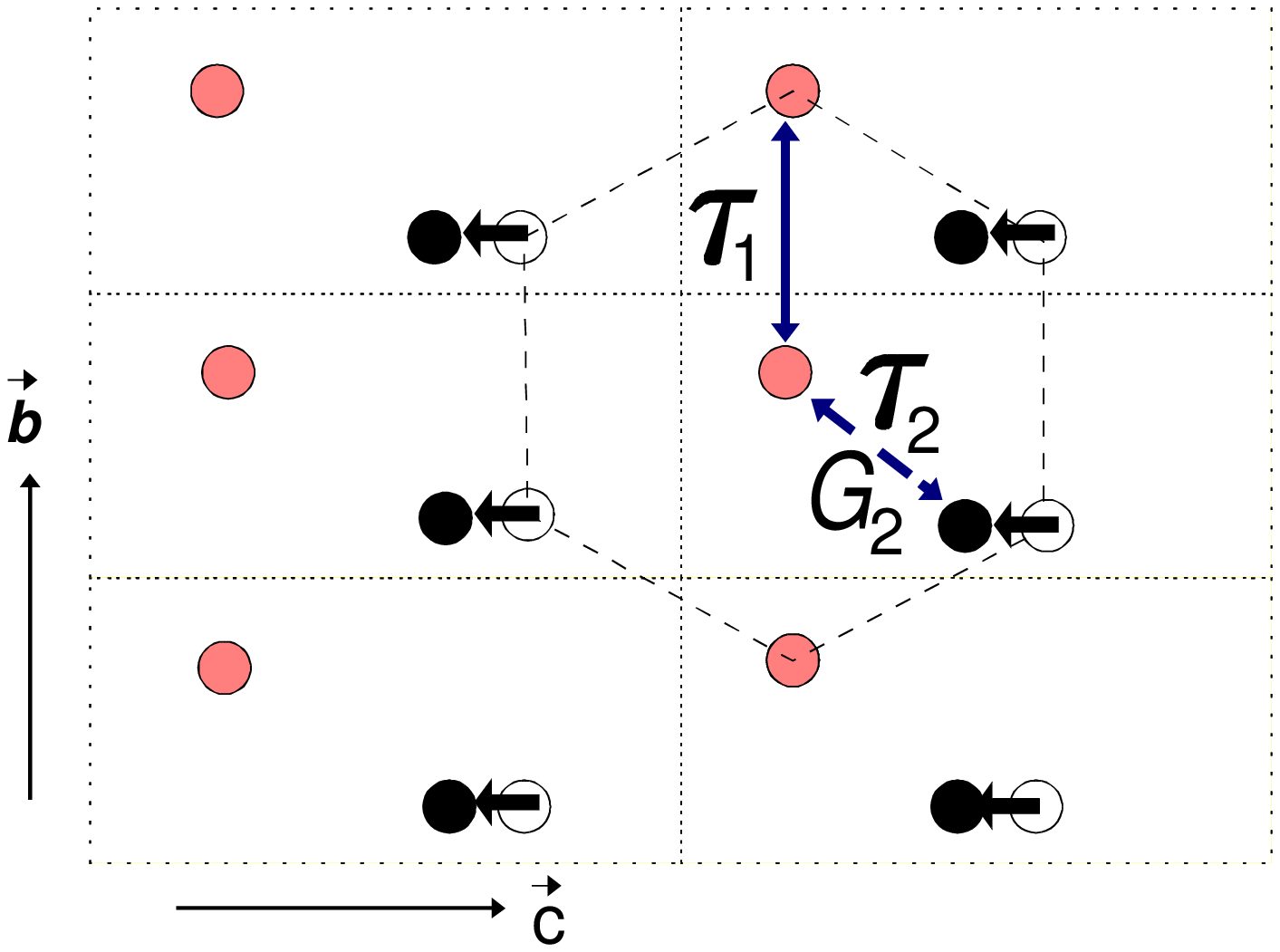}
\leavevmode
\mbox{} \\ \mbox{} \\
Fig.\ 4.
\end{figure}

\pagebreak

\begin{figure}[e]
\epsfxsize=16cm
\epsfysize=12cm
\epsfbox{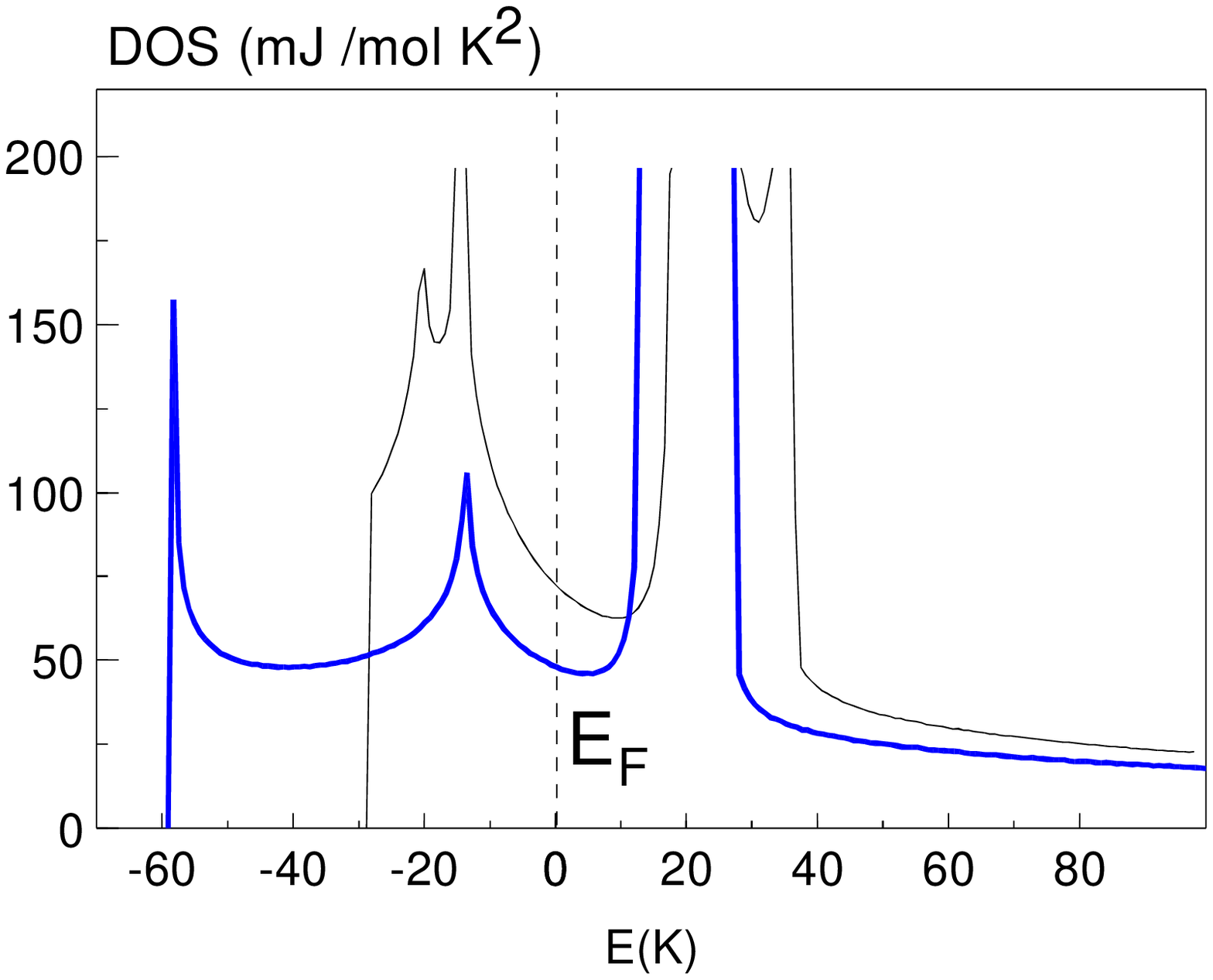}
\leavevmode
\mbox{} \\ \mbox{} \\
Fig.\ 5.
\end{figure}

\pagebreak

\begin{figure}[e]
\epsfxsize=16cm
\epsfysize=12cm
\epsfbox{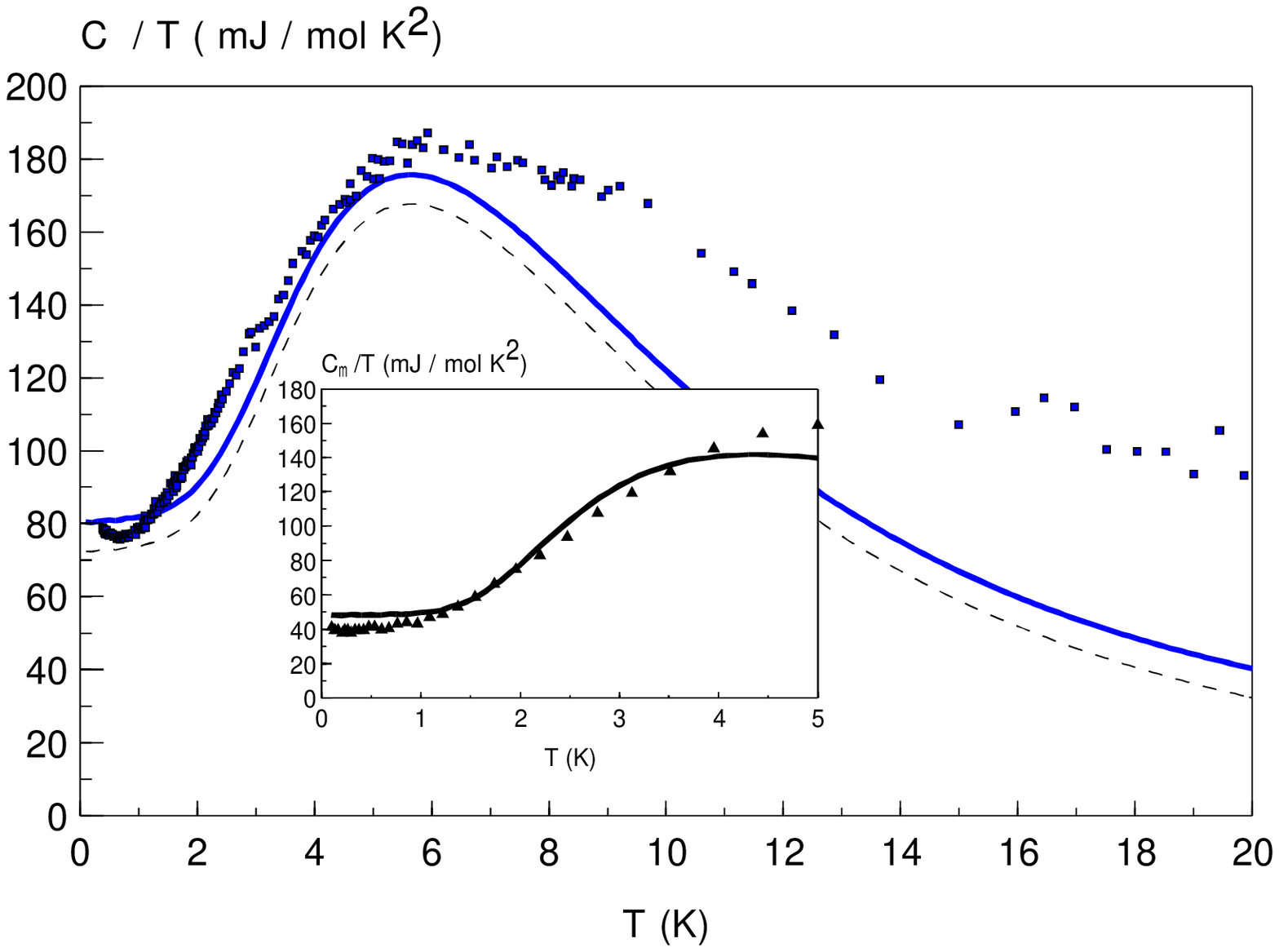}
\leavevmode
\mbox{} \\ \mbox{} \\
Fig.\ 6.
\end{figure}

\pagebreak

\begin{figure}[e]
\epsfxsize=16cm
\epsfysize=12cm
\epsfbox{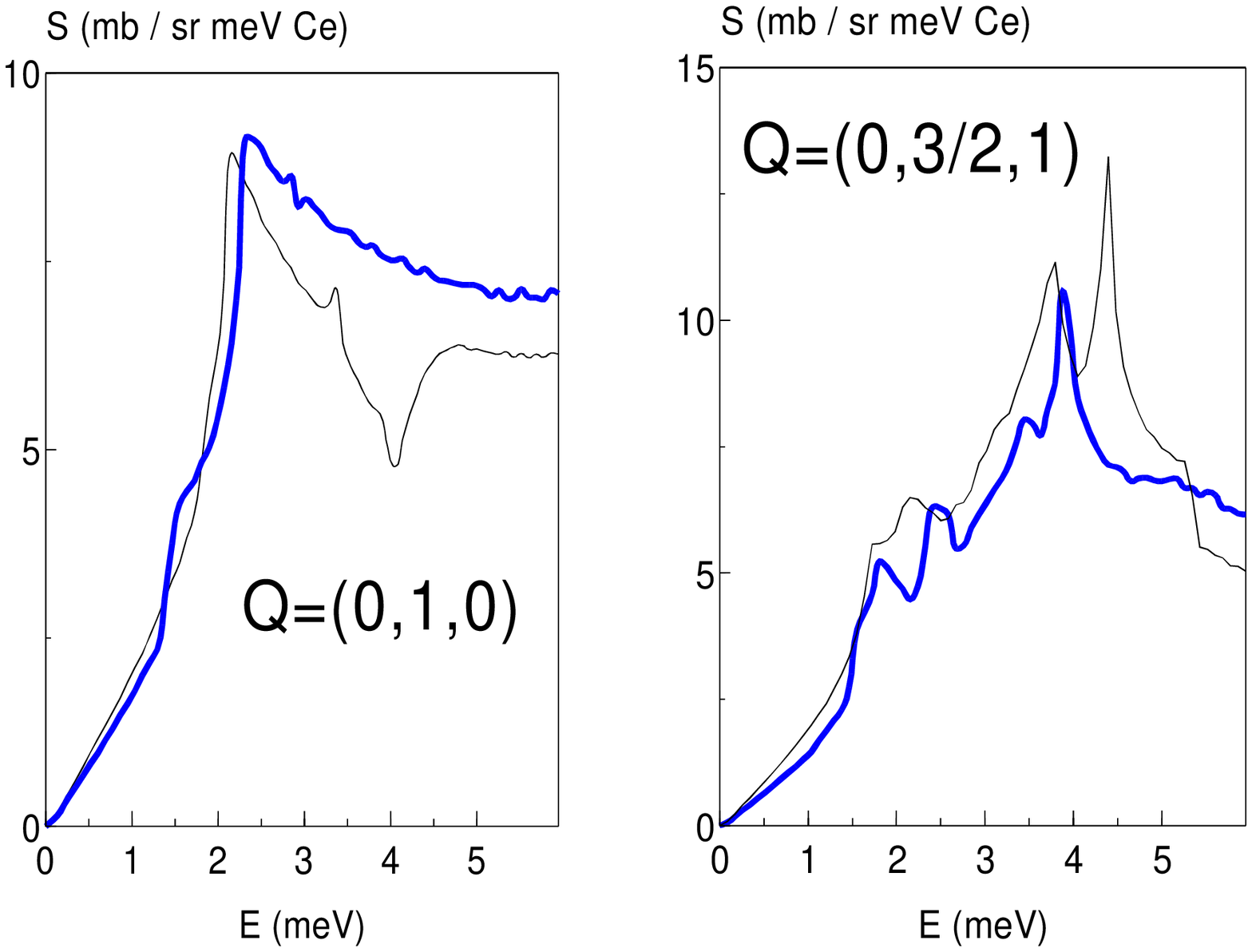}
\leavevmode
\mbox{} \\ \mbox{} \\
Fig.\ 7.
\end{figure}

\pagebreak

\begin{figure}[e]
\epsfxsize=16cm
\epsfysize=12cm
\epsfbox{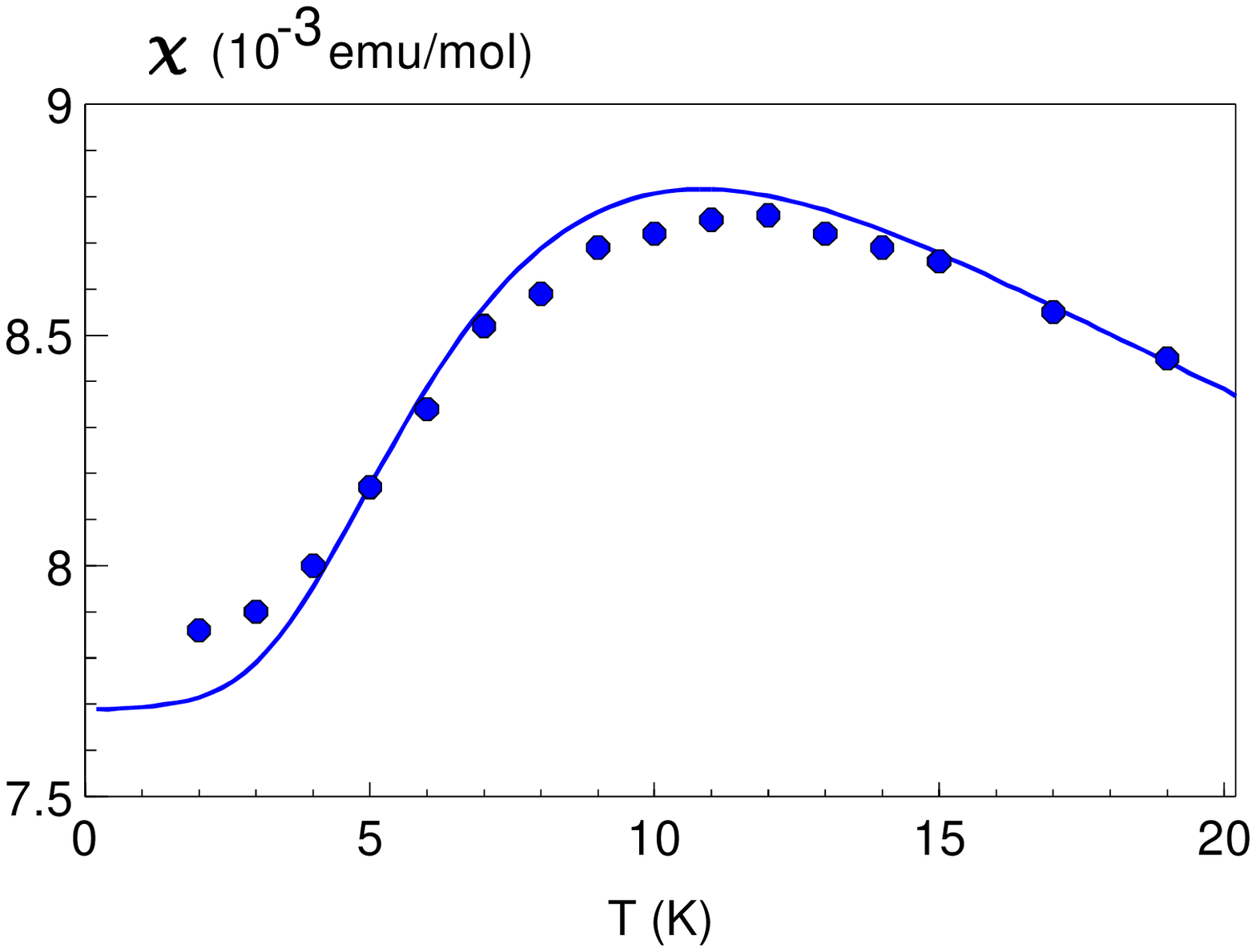}
\leavevmode
\mbox{} \\ \mbox{} \\
Fig.\ 8.
\end{figure}

\pagebreak

\begin{figure}[e]
\epsfxsize=16cm
\epsfysize=12cm
\epsfbox{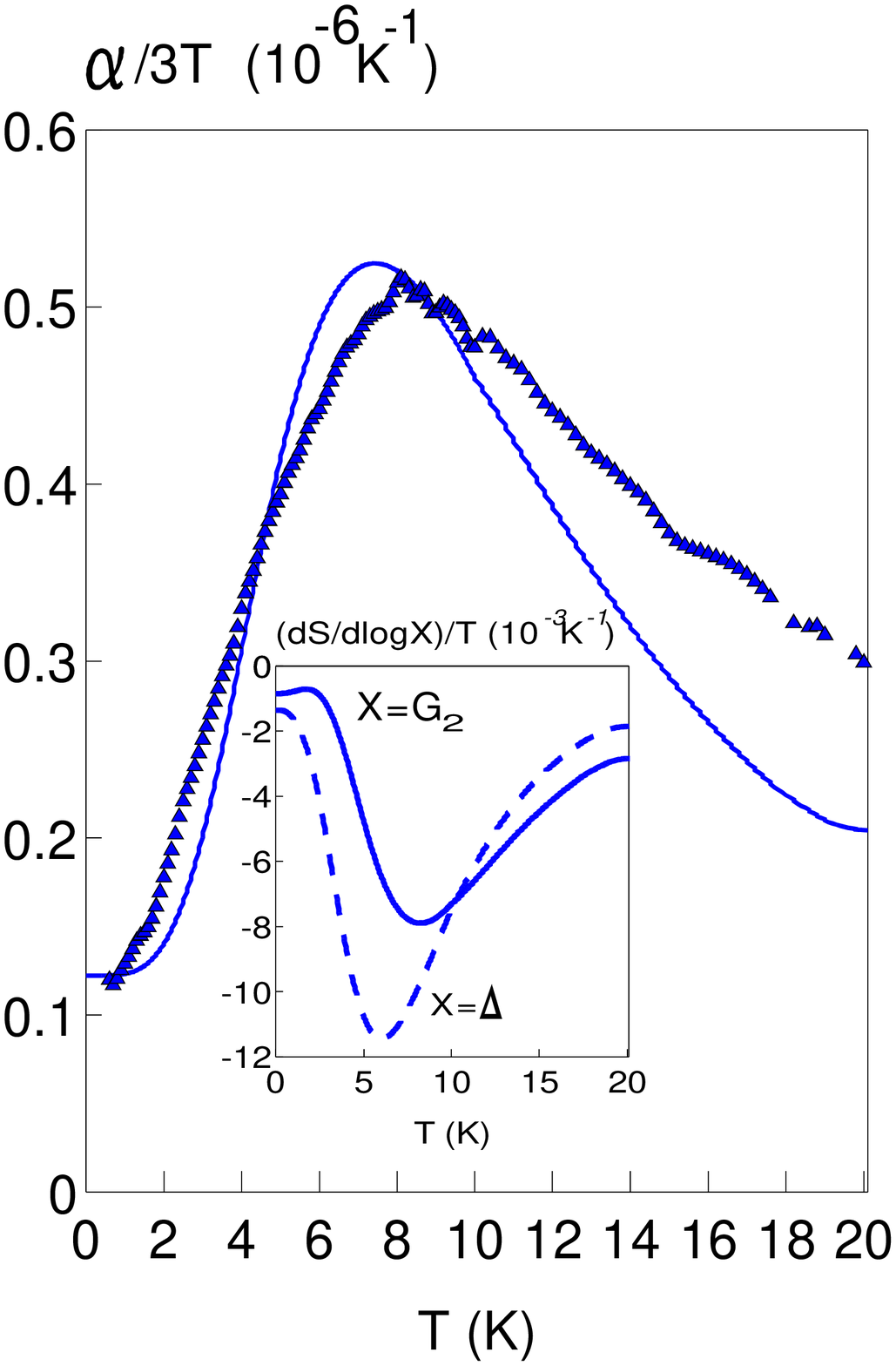}
\leavevmode
\mbox{} \\ \mbox{} \\
Fig.\ 9.
\end{figure}

\pagebreak

\begin{figure}[e]
\epsfxsize=12cm
\epsfysize=15cm
\epsfbox{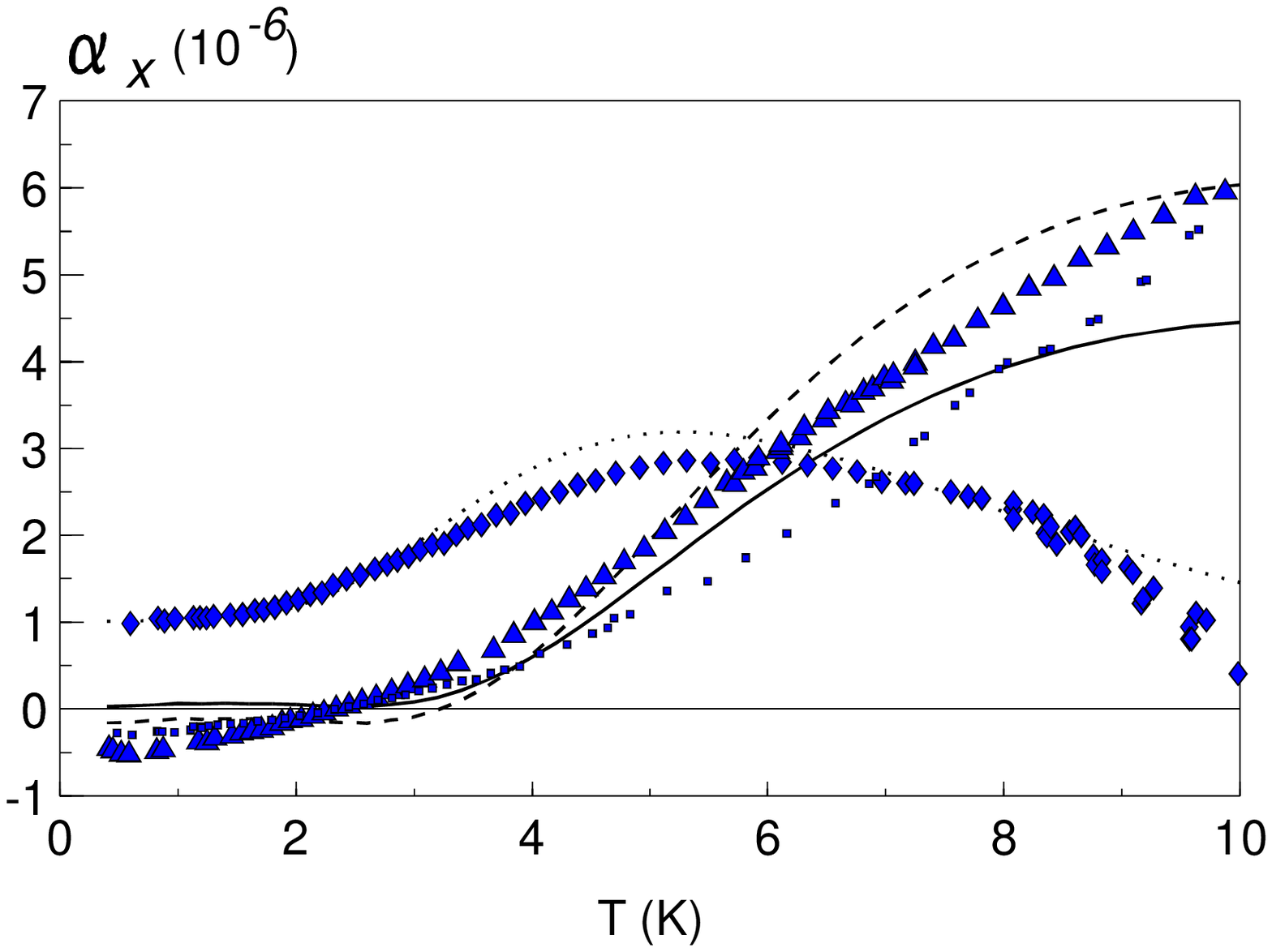}
\leavevmode
\mbox{} \\ \mbox{} \\
Fig.\ 10.
\end{figure}

\pagebreak

\begin{figure}[e]
\epsfxsize=12cm
\epsfysize=17cm
\epsfbox{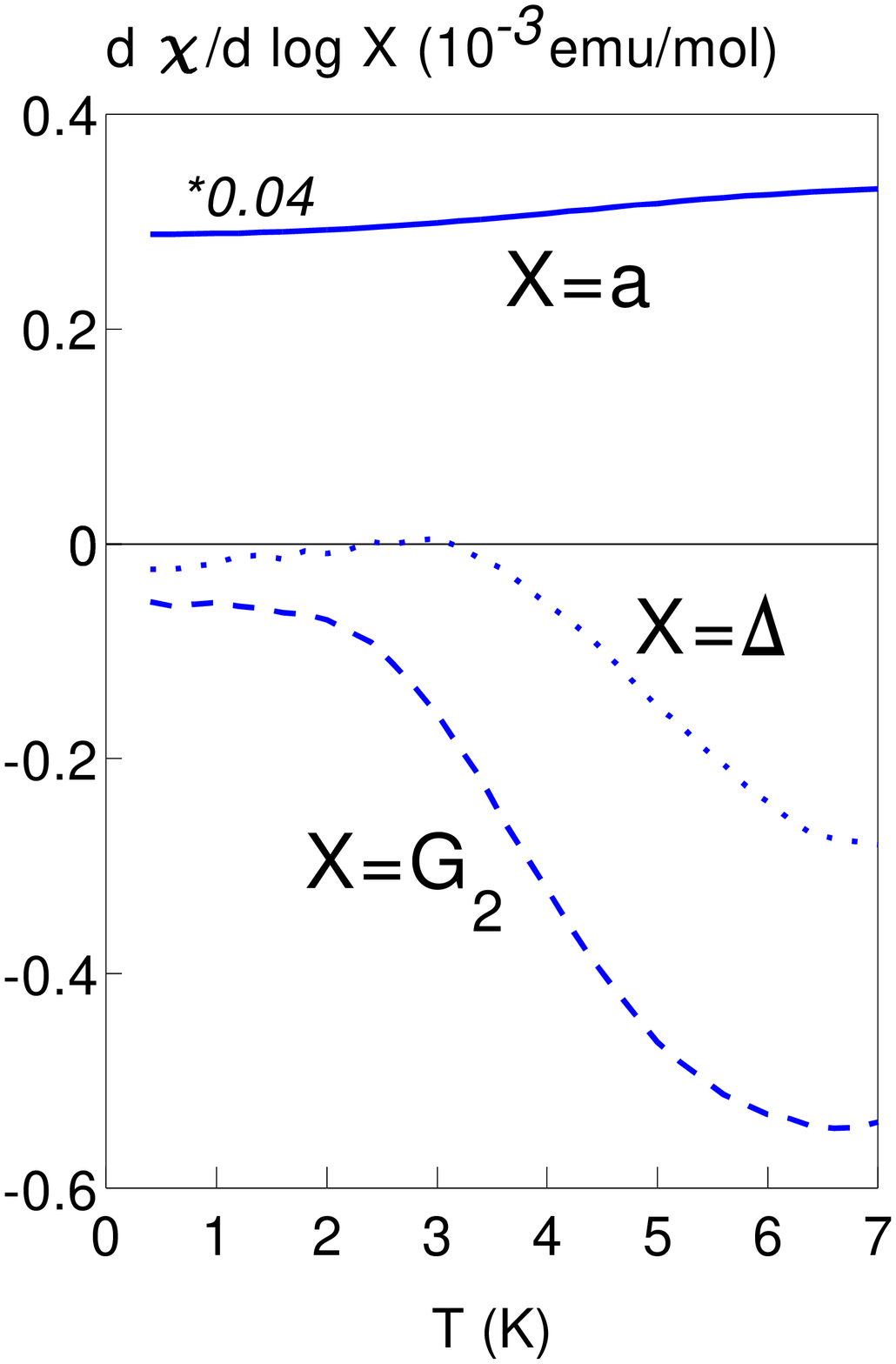}
\leavevmode
\mbox{} \\ \mbox{} \\
Fig.\ 11.
\end{figure}

\pagebreak

\begin{figure}[e]
\epsfxsize=14cm
\epsfysize=12cm
\epsfbox{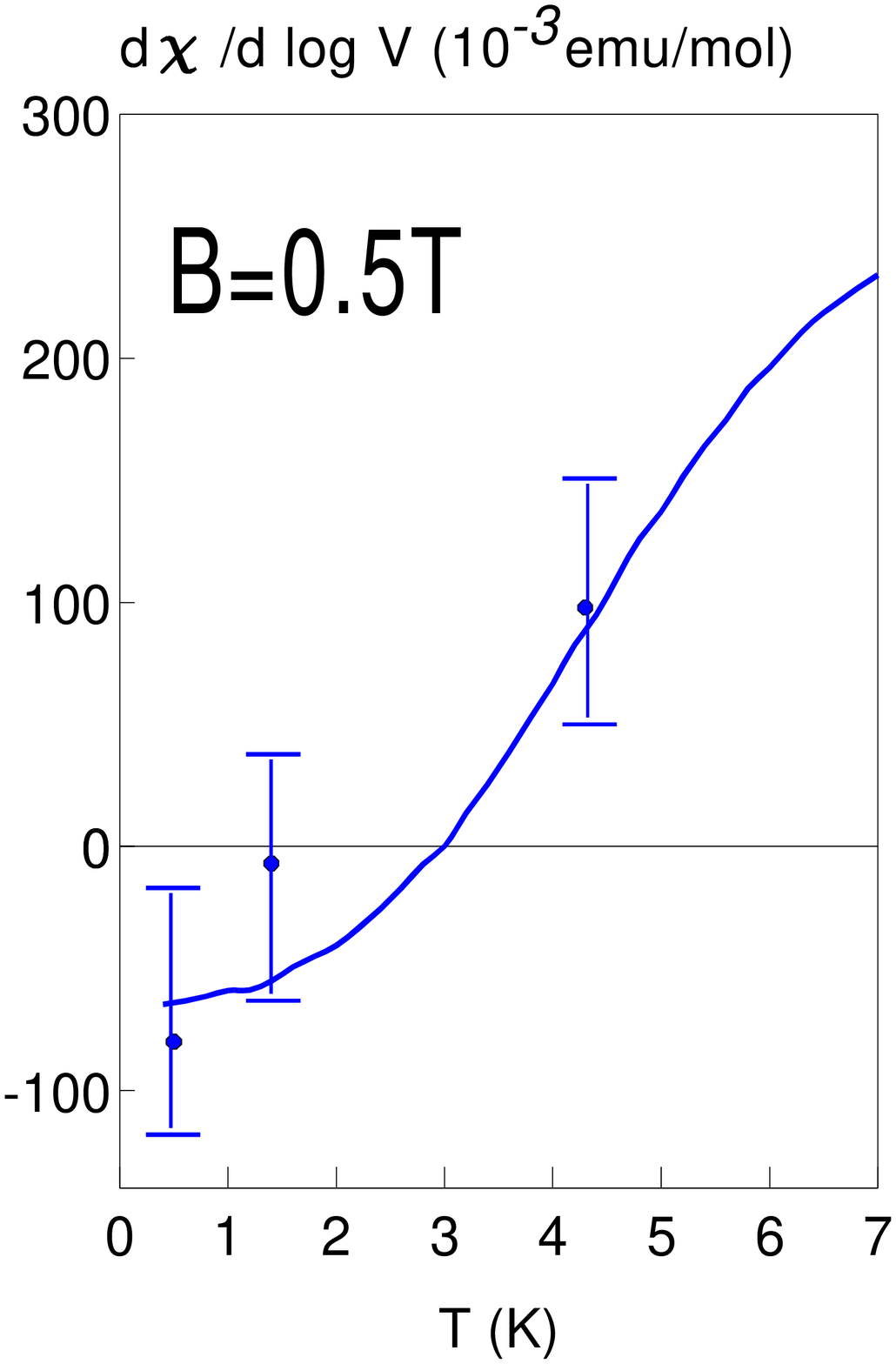}
\leavevmode
\mbox{} \\ \mbox{} \\
Fig.\ 12.
\end{figure}

\end{document}